\documentclass[aps,jcp,superscriptaddress,amsmath,amssymb]{revtex4-1}

\usepackage{comment}
\usepackage{tikz}
\usetikzlibrary{trees}
\usetikzlibrary{decorations.pathmorphing}
\usetikzlibrary{decorations.markings}
\tikzset{
    photon/.style={decorate, decoration={snake,segment length=1.5mm}, draw=black},
    coulomb/.style={dotted},
    electron/.style={draw=black, postaction={decorate},
        decoration={markings,mark=at position .55 with {\arrow[draw=black]{>}}}}, 
    gluon/.style={decorate, draw=magenta,
        decoration={coil,amplitude=4pt, segment length=5pt}},
    boundelectron/.style={thick, double},
    transverse/.style={dashed}
}

\usepackage{graphicx}
\usepackage{dcolumn}
\usepackage{bm}
\usepackage{rotating}

\newcolumntype{.}{D{.}{.}{8}}

\usepackage[utf8]{inputenc}
\usepackage{physics}
\usepackage{amsmath, amssymb}
\usepackage{tabularx,booktabs,array,dcolumn}
\usepackage{setspace}
\usepackage{vmargin}
\usepackage{xcolor}
\usepackage{graphicx,psfrag,subfigure}

\usepackage{float}
\usepackage[all]{xy}

\usepackage{bm}
\usepackage{mathtools}
\usepackage{physics}
\usepackage[english]{babel}

\newcommand{\bos}[1]{\boldsymbol{#1}}
\newcommand{\mr}[1]{\mathrm{#1}}
\newcommand{\pd}[2]{\frac{\partial #1}{\partial #2}}

\def\Eh{\text{E}_\text{h}}

\def\iim{\mr{i}}
\def\eem{\mr{e}}

\def\DC{\text{DC}}
\def\DCB{\text{DCB}}

\def\np{N} 
\def\nnuc{N_\text{nuc}} 
\def\nb{N_\text{b}} 

\def\proj{\text{proj}}
\def\unittwo{1^{[2]}} 
\def\unitfour{1^{[4]}} 

\def\four{^{[4]}} 

\def\bp{\bos{p}}
\def\br{\bos{r}}
\def\bs{\bos{s}}
\def\bk{\bos{k}}
\def\balpha{\bos{\alpha}}
\def\bsigma{\bos{\sigma}}

\def\bA{\mathbf{A}}
\def\bE{\bos{E}}
\def\be{\boldsymbol{e}}
\def\intdr{\int \text{d} \br}

\usepackage[unicode]{hyperref}
\usepackage{soul}

\usepackage[unicode]{hyperref}
\hypersetup{
   unicode=true,          
   plainpages=false,
   colorlinks=true,       
   citecolor=blue,        
}

\usepackage{url}

\begin{document}

\title{%
On the Breit interaction in an explicitly correlated variational Dirac--Coulomb framework
}

\author{D\'avid Ferenc} 
\author{P\'eter Jeszenszki} 
\author{Edit M\'atyus} 
\email{edit.matyus@ttk.elte.hu}
\affiliation{ELTE, Eötvös Loránd University, Institute of Chemistry, 
Pázmány Péter sétány 1/A, Budapest, H-1117, Hungary}

\date{\today}

\begin{abstract}
\noindent %
The Breit interaction is implemented in the no-pair variational Dirac--Coulomb (DC) framework 
using an explicitly correlated Gaussian basis reported in the previous paper [P.~Jeszenszki, D.~Ferenc, and E.~Mátyus (2022)].
Both a perturbative and a fully variational inclusion of the Breit term is considered. 
The no-pair DC plus perturbative Breit as well as the no-pair Dirac--Coulomb--Breit (DCB) energies are compared with perturbation theory results including the Breit--Pauli Hamiltonian and leading-order non-radiative quantum electrodynamics corrections for low $Z$ values.
Possible reasons for the observed deviations are discussed.
\end{abstract}

\maketitle
\clearpage
\noindent %
\section{Introduction \label{ch:intro}}
\noindent %
Relativistic effects play an essential role in the quantitative understanding of high-resolution atomic and molecular spectra. For atoms and molecules with a low $Z$ nuclear charge number, the non-relativistic wave function provides an adequate zeroth-order approximation, and relativistic effects can be treated as perturbation. 
The most common route for the theoretical determination of energy levels of low-$Z$ atoms and molecules is provided by the non-relativistic quantum electrodynamics (nrQED) framework, in which the leading-order relativistic corrections are the well-known Breit--Pauli Hamiltonian terms. The nrQED approach gives excellent agreement with high-resolution spectroscopy measurements for several atomic and molecular systems \cite{test21,AlGiCoKoSc20,FeKoMa20,PuKoCzPa16}. 
At the same time, the derivation of the correction operators is tedious, one has to deal with (cancellation of) divergent terms at higher orders \cite{Pa06,PuKoCzPa16,PaYeVlPa21}. For larger nuclear charge numbers, this approach becomes unsuitable to reach high theoretical accuracy. 

An alternative approach is provided by the variational solution of the Dirac--Coulomb (DC) or Dirac--Coulomb--Breit (DCB) equation which also bears several difficulties. The main complication is caused by the coupling of the positive and negative energy states by the electron-electron interaction known as continuum dissolution or the Brown--Ravenhall disease (BR) \cite{BrRa51}. The solution to this problem was derived from quantum electrodynamics (QED). Sucher proposed the projection of the Dirac operator to the positive energy ($E_+$) subspace of some non-interacting reference problem \cite{sucherPhD1958,Su80}.
This approach, during which the electron-positron pair contribution is eliminated from the Hamiltonian, is referred to as the `no-virtual-pair' approximation, and it is commonly used in modern relativistic quantum chemistry computations.

There are various options for choosing an (effective) one-particle reference problem to define a positive-energy projector. 
Mittleman argued for a Hartree--Fock (HF) based projector \cite{Mittleman1981} and methods with a single reference determinant are widely used in computations \cite{dirac2020, PaFFGr96, QuSkGr98}. 
Liu and co-workers pointed out that if the projector is defined with the orbitals of the HF computation, then the positive-energy projected Hamiltonian (and the QED pair corrections to it \cite{LiLi12}) can be naturally written in a second-quantized form \cite{Liu12,Liu14} using the same orbitals as a basis. 
More recently, Almoukhalalati \emph{et al.} considered the choice of a `good' projector for computations including electron correlation \cite{AlKnJeDySa2016}. 
Liu and co-workers \cite{LiShLi12} noted that combination of the orbital-based formalism and explicit correlation (`F12 methods') is non-trivial, and proposed a `dual basis' approach.

Over the past two decades, methods using explicitly correlated basis functions have been developed and used for solving the DC equation of (helium-like) atoms \cite{pestka_variational_1998,pestka_application_2006,pestka_dirac-coulomb_2003,SiMaRe15}. 
Explicitly correlated, non-separable basis functions represent a departure from the single-particle picture, and the construction of an $E_+$ projection operator is not immediately obvious in this framework. 
For atomic computations, Bylicki, Pestka, and Karwowski proposed to use
the complex coordinate rotation (CCR) technique to separate the $E_+$ subspace of the non-interacting problem in the explicitly correlated basis, and employed this projection technique to compute the no-pair Dirac--Coulomb energy for the ground state of helium-like ions (atom) with Hylleraas-type basis functions
\cite{ByPeKa08,pestka_dirac-coulomb_2003,PeByKa07}.

We have adapted this projector for solving the Dirac--Coulomb and Dirac--Coulomb--Breit models using explicitly correlated Gaussian (ECG) basis sets for atoms and also for molecules. A short account of this work was first given in Ref.~\cite{JeFeMa21}. The Dirac--Coulomb implementation is reported in detail in Ref.~\cite{JeFeMa21dc} (henceforth Paper~I), and the present work reports theoretical, algorithmic, and numerical details regarding the implementation of the Breit term in the explicitly correlated DC framework of Paper~I. 
No-pair energies computed in this work are reported for the helium atom and for small and light molecules converged to high precision and the results are compared with energies computed in the nrQED framework.

Throughout this work, Hartree atomic units are used and the speed of light is $c=\alpha^{-1}a_0\Eh/\hbar$ with $\alpha^{-1}=137.$035 999 084  \cite{codata18}.

\subsection{The Breit interaction}
\begin{figure}[H]
    \begin{center}
    \begin{tikzpicture}
    
    \draw[boundelectron] (1.5,.0) -- (2.5,0) ;
    \draw[boundelectron] (1.5,-1.0) -- (2.5,-1) ;
    \draw[photon,thick] (2.0,-1) -- (2.0,0) ;
    
    \node at (3.5,-.5) {$=$};
    
    \draw[boundelectron] (4.0,.0) -- (5,0) ;
    \draw[boundelectron] (4.0,-1.0) -- (5,-1) ;
    \draw[coulomb,thick] (4.5,-1) -- (4.5,0) ;

    \node at (5.5,-.5) {$+$};

    \draw[boundelectron] (6,.0) -- (7,0) ;
    \draw[boundelectron] (6,-1.0) -- (7,-1) ;
    \draw[transverse,thick] (6.5,-1) -- (6.5,0) ;
    
    \end{tikzpicture} 
    \end{center}
    \caption{The single-photon exchange diagram in the Coulomb gauge can be separated into an instantaneous Coulomb- and a retarded transverse photon exchange contribution.}
    \label{fig:onephoton}
\end{figure}
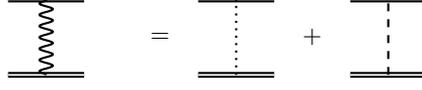

The classical Lagrangian and Hamiltonian dynamics of moving charges was studied by Darwin \cite{Darwin1920} to order $\mathcal{O}(1/c^2)$. Breit identified the velocity operator as $c \bos{\alpha}$ in relativistic quantum mechanics \cite{Breit1928} and replaced the classical velocity in Darwin's expression to arrive at the quantum mechanical form for the interaction of two moving charges \cite{Breit1929}.
A more consistent approach is provided by the derivation of an effective potential from QED scattering amplitudes \cite{InMo2016}. 
The tree-level diagram in the $S$ matrix of an electron scattering process is the one-photon exchange diagram (Fig.~\ref{fig:onephoton}). 
The $S$ matrix is related to the shift in the energy levels by the Gell-Mann--Low--Sucher formula \cite{GMLo51,Su57} that can be used to calculate the effect of the interaction with the radiation field on the energy levels. The external field of the nuclei can be taken into account exactly within the Furry picture \cite{Fu51}.

The Coulomb gauge, for which the vector potential has a vanishing divergence, is a natural and convenient choice for describing the molecular domain, since, in addition, to the instantaneous Coulomb potential, there are only transverse photons present (Fig.~\ref{fig:onephoton}).
The photon propagator in the Coulomb gauge is
\begin{align}
    &D_{00}^{\text{C}} (\br_{12}) = \frac{1}{4\pi r_{12}} \nonumber \\ 
    & D_{0i}^{\text{C}} = D_{i0}^{\text{C}}=0 \nonumber  \\
    &D_{ij}^{\text{C}} (\br_{12},\omega/c) =  \int \frac{d^3\bk}{(2\pi)^3} \frac{\eem^{\iim\bk\cdot \br_{12}}}{\omega^2/c^2-\bk^2+\iim\varepsilon} \left(\delta_{ij} - \frac{k_i k_j}{\bk^2}\right) \; ,
    \label{eq:Ddef}
\end{align}
and the corresponding one-photon exchange interaction is (for example, p.~72 of Ref.~\cite{lindgrenRelativisticManyBodyTheory2011})
\begin{align}
    I^\text{C}(\omega/c)
    =
    q_1q_2
    \left[%
      \frac{1}{r_{12}}
      -
      \balpha_1\cdot\balpha_2\frac{\cos \left( \omega r_{12}/c\right)}{r_{12}}
      +
      \left\lbrace%
        (\balpha_1\cdot\grad_1)(\balpha_2\cdot\grad_2)\frac{\cos \left( \omega r_{12}/c\right)-1}{\omega^2r_{12}/c^2} 
      \right\rbrace
    \right] \; ,
    \label{eq:onephx}
\end{align}
where $\omega$ is the frequency of the exchanged photon and the gradient operators act only within the braces.  
We may expand Eq.~(\ref{eq:onephx}) in terms of $\omega r_{12}/c$  (for example, p.~253 of Ref.~\cite{ReWoBook15})
\begin{align}
  \frac{1}{q_1q_2} I^\text{C}(\omega/c)
  &=
  \frac{1}{r_{12}}
  -
  \frac{\balpha_1\cdot\balpha_2}{r_{12}}
  \left[%
    1-\frac{1}{2}\left(\frac{\omega r_{12}}{c}\right)^2 +\ldots 
  \right]
  \nonumber \\
  &\ +\left\lbrace%
    (\balpha_1\cdot\grad_1)(\balpha_2\cdot\grad_2)
    \left[%
      - \frac{1}{2}\left(\frac{\omega r_{12}}{c}\right)^2 
      + \frac{1}{4!}\left(\frac{\omega r_{12}}{c}\right)^4 - \ldots
    \right]
    \frac{c^2}{\omega^2r_{12}} 
  \right\rbrace
  \nonumber \\
  &=
  \frac{1}{r_{12}}
  -
  \balpha_1\cdot\balpha_2\left(\frac{1}{r_{12}} +\mathcal{O}(c^{-2})  \right) \nonumber \\
  &\ +
  \left\lbrace%
    (\balpha_1\cdot\grad_1)(\balpha_2\cdot\grad_2) \left(-\frac{1}{2}r_{12}\right) + \mathcal{O}(c^{-2})
  \right\rbrace    
  \nonumber \\
  &=
  \frac{1}{r_{12}}
  -
  \frac{\balpha_1\cdot\balpha_2}{r_{12}}
  -\frac{1}{2} \left\lbrace (\balpha_1\cdot\grad_1)(\balpha_2\cdot\grad_2) r_{12} \right\rbrace 
  +\mathcal{O}(c^{-2}) \; .
\end{align}
If we neglect the $\mathcal{O}(c^{-2})$ terms, which is often called the zero-frequency or non-retardation approximation, we obtain the sum of the Coulomb and the Breit terms
\begin{align}
    I^\text{C}(0)  
    &=
    \frac{q_1q_2}{r_{12}}+B(1,2) \\
    B(1,2)
    &=
    -q_1q_2
    \left[%
      \frac{\balpha_1\cdot\balpha_2}{r_{12}}
      +
      \frac{1}{2}\left\lbrace(\balpha_1\cdot\grad_1)(\balpha_2\cdot\grad_2)r_{12}\right\rbrace
    \right] \; .
    \label{eq:Breit1}
\end{align}
If the differentiation is carried out (similarly to the earlier notation, the differential operators act only within the braces), we obtain the more common form for the Breit operator
\begin{align}
    B(1,2)=-\frac{q_1q_2}{2}\left[\frac{\balpha_1\cdot\balpha_2}{r_{12}}+\frac{(\balpha_1\cdot\bos{r}_{12})(\balpha_2\cdot\bos{r}_{12})}{r_{12}^3}\right] \;.
    \label{eq:Breit2}
\end{align}
It is interesting to note that both the Coulomb and the Breit interactions arise from the `zero-frequency approximation' of the exact one-photon exchange, and this feature suggests that they should be treated on an equal footing, \emph{e.g.,} in a variational procedure. 

There has been a long discussion in the literature about the Breit term whether it can be included in a variational procedure or a perturbative treatment should be preferred.
It has been argued by Bethe and Salpeter in 1957 \cite{BeSabook57} that the Breit interaction should only be used in first-order perturbation theory, since the coupling with negative-energy intermediate states would yield second- and higher-order corrections that are too large. 
Later, Sucher in 1980 \cite{Su80} pointed out that if the operators are defined with positive energy projection, then the inclusion of the Breit interaction on the same footing as the Coulomb interaction in a variational scheme is appropriate.
The discussion continues also in the numerical quantum chemistry literature based on formal as well as practical considerations and observations \cite{Kim67,Kagawa75,QuGrWi87,LiMPYnOs89,ThSc10}.

In the first order of perturbation theory, the Breit interaction corresponds to a single transverse photon exchange, meanwhile at higher orders, it describes the consecutive exchange of several transverse photons. 
If it is included in a variational computation, it accounts for a sum (`ladder') of all-, \emph{i.e.,} one-, two-, three-, etc. non-retarded transverse photon exchanges.

\section{The no-pair Dirac--Coulomb--Breit equation \label{ch:npDCB}}
\noindent In Paper~I \cite{JeFeMa21dc}, we have described in detail a (quasi-)variational procedure for the solution of the no-pair Dirac equation. 
The no-pair Dirac operator for $N$ spin-1/2 particles is
\begin{align}
  {H}
  =
  \sum_{i=1}^N
  \Lambda_+ (h_{i}^{[4^\np]}+ u_{i}1^{[4^\np]}) \Lambda_+
  + 
  \sum_{i=1}^N \sum_{j>i}^N
  \Lambda_+ v_{ij} \Lambda_+  \; ,
  \label{eq:dirac}
\end{align}
where $h_{i}^{[4^\np]}=\unitfour(1)\boxtimes \ldots \boxtimes h\four_i(i) \boxtimes \ldots \boxtimes \unitfour(N)$
with $h\four_i= c \balpha\four \cdot \bp + \beta\four m_i c^2$ is the free, one-particle Dirac operator; $u_i=\sum_{a=1}^{\nnuc}q_iQ_a/|r_i-R_a|$ is the Coulomb interaction energy with the fixed, point-like nuclei; $\Lambda_+$ is the $E_+$-projection operator, and $v_{ij}$ is either the Coulomb or the Coulomb--Breit interaction.  We use the block-wise direct product, also called Tracy--Singh product \cite{LiShLi12}, for convenience.

The many-particle Coulomb and Breit operators have the following form
\begin{align}
    v_{ij}^{[4^\np]}&=\frac{1}{r_{ij}} 1^{[4^\np]} + \sum_{k=1}^3\sum_{l=1}^3\left(b_{ij}\right)_{kl}\left[ \unitfour (1) \boxtimes \dots \left(\balpha (i)\right)_k \boxtimes \dots \boxtimes \left(\balpha (j)\right)_l \boxtimes \dots \boxtimes \unitfour (\np) \right] \\
    \left(b_{ij}\right)_{kl}&=-\frac{1}{2}\left(  \frac{\delta_{kl}}{r_{ij}}+\frac{(\br_{ij})_k (\br_{ij})_l}{r_{ij}^3} \right) \; .
\end{align}
The explicit matrix form of the no-pair Dirac--Coulomb--Breit Hamiltonian for two particles is
\begin{align}
  &H(1,2) = \nonumber \\
  &{ \footnotesize
  \Lambda_+
  \left(%
    \begin{array}{@{} c@{}c@{}c@{}c @{}}
       V\unitfour+U \unitfour & 
       c \bsigma\four_2 \cdot \bp_2 & 
       c \bsigma\four_1 \cdot \bp_1 & 
       B \\
       c\bsigma\four_2 \cdot \bp_2 & 
       V\unitfour+(U - 2m_2c^2)\unitfour & 
       B & 
       c \bsigma\four_1 \cdot \bp_1 \\
       c\bsigma\four_1 \cdot\bp_1 & 
       B &
       V\unitfour+(U-2m_1c^2)\unitfour & 
       c \bsigma\four_2 \cdot \bp_2 \\
       B & 
       c \bsigma\four_1 \cdot \bp_1 &
       c \bsigma\four_2 \cdot \bp_2 & 
       V\unitfour+(U-2m_{12}c^2)\unitfour \\
    \end{array}
  \right)
  \Lambda_+
  }
  \label{eq:fullHam}
\end{align}
with $m_{12}=m_1+m_2$, $\bp_i = -\iim(\pd{}{r_{ix}},\pd{}{r_{iy}},\pd{}{r_{iz}})$ ($i=1,2$),
$\bsigma\four_1=(\sigma_x\otimes\unittwo,\sigma_y\otimes\unittwo,\sigma_z\otimes\unittwo)$
and
$\bsigma\four_2=(\unittwo\otimes \sigma_x,\unittwo\otimes\sigma_y,\unittwo\otimes\sigma_z)$, where $\sigma_x,\sigma_y,$ and $\sigma_z$ are the $2\times 2$ Pauli matrices, and $U=\sum_{i=1}^n \sum_{a=1}^{\nnuc}q_iQ_a/|\bos{r}_i-\bos{R}_a|$ is the external Coulomb potential of the nuclei. 
We note that a $-2m_ic^2$ shift is introduced for both particles in the operator in Eq.~(\ref{eq:fullHam}) in comparison with Eq.~(\ref{eq:dirac}) that is a commonly used energy-scale shift and serves here practical, computational purposes.

The $\Lambda_+$ projection operator is constructed from the positive-energy solutions of the reference problem, which is chosen to be the external-field Dirac equation without electron-electron interaction,
\begin{align}
    \sum_{i=1}^\np\left( h_i^{[4^\np]}+u_i 1^{[4^\np]} \right)\phi_k=\epsilon_k \phi_k && \Lambda_+=\sum_{k\in\lbrace E_+\rbrace} \ket{\phi_k}\bra{\phi_k} \; .
    \label{eq:nonintproj}
\end{align}
The projected Hamiltonian is bounded from below, and it has well-defined bound states, free of the BR problem. 
Technical details regarding the projection techniques are explained in Sec.~\ref{ch:proj} and in Paper~I \cite{JeFeMa21dc}.

In Eq.~(\ref{eq:fullHam}), the $B$ blocks along the anti-diagonal of the matrix represent the Breit potential
\begin{align}
    B
    =
    G
    -
    \frac{q_1q_2}{2}
    \sum_{i=1}^3\sum_{j=1}^3
      \sigma^{[4]}_{1_i} \sigma^{[4]}_{2_j} 
        \left\lbrace \grad_{1_i}\grad_{2_j}r_{12} \right\rbrace \ .
    \label{eq:breit}
\end{align}
The first term of $B$ is called the Gaunt interaction, which reads for two particles as 
\begin{align}
     G=-\frac{q_1q_2}{r_{12}}\bsigma^{[4]}_1\cdot\bsigma^{[4]}_2
     =-\frac{q_1q_2}{r_{12}}\begin{pmatrix*}[c]
     1 & 0 & 0 & 0 \\
     0 &-1 & 2 & 0 \\
     0 & 2 &-1 & 0 \\
     0 & 0 & 0 & 1
    \end{pmatrix*}   \; .
    \label{eq:gaunt}
\end{align}
The wave function is expanded over a set of sixteen-component ($2^N$ for $N$ particles) basis spinors $\bos{d}_\chi$ constructed using $\nb$ explicitly correlated Gaussian functions
\begin{align}
     \Psi(\br_1,\br_2) &=\mathcal{A} \sum_{i=1}^{\nb} \sum_{\chi=1}^{16} c_{i\chi} \bos{d}_{\chi}
     \Theta_i\left(\br_1,\br_2;\bos{A}_i,\bos{s}_i\right) 
     \label{eq:spinorbasis} \\
     \Theta_i\left(\br_1,\br_2;\bos{A}_i,\bos{s}_i\right) &= \exp\left[ -(\br-\bos{s}_i)^T\left(\bos{A}_i\otimes 1^{[3]}\right)(\br-\bos{s}_i)\right]  \; ,
     \label{eq:ECGbasis}
\end{align}
where $\br=(\br_1,\br_2)^T$ are the coordinates of the particles (electrons), $\bos{s}_i\in \mathbb{R}^{6}$ and $\bos{A}_i \in \mathbb{R}^{2 \times 2} $ are parameters of the basis functions, and $\mathcal{A}$ is the anti-symmetrization operator providing the proper permutational symmetry for the relativistic two-fermion wave function \cite{JeFeMa21dc}.

\subsection{Kinetic balance}
\label{ssec:KB}
The large and small components of a relativistic four-component basis function must satisfy the relation
\begin{align}
    \psi^\text{s}=\frac{\bsigma^{[2]} \bp }{2mc} \psi^\text{l} \; 
    \label{eq:kb}
\end{align}
following from the kinetic balance condition as discussed in Refs.~\cite{Ku84,Liu10}.
We use here the simplest two-particle generalization of the one-electron kinetic balance condition and implement the two-electron kinetic balance condition in the sense of a transformation or metric \cite{Ku84}:
\begin{align}
    H_\text{KB}=X^\dagger H X \; , 
    && 
    X=\text{diag}\left(%
      1\four, \frac{\left(\bsigma_2\four \bp_2 \right)}{2m_2c},%
      \frac{\left(\bsigma_1\four \bp_1 \right)}{2m_1c},%
      \frac{\left(\bsigma_1\four \bp_1 \right)\left(\bsigma_2\four \bp_2 \right)}{4m_1m_2c^2}\right) \; .
    \label{eq:XHX} 
\end{align}
The two-particle wave function, 
which corresponds to the block-wise direct product form of the 2-electron operators, has the following structure 
\begin{align}
    \bos{\Psi}(\br_1,\br_2)=\left( \begin{array}{c}
         \psi^\text{ll}(\br_1,\br_2)  \\
         \psi^\text{ls}(\br_1,\br_2)  \\
         \psi^\text{sl}(\br_1,\br_2)  \\
         \psi^\text{ss}(\br_1,\br_2)  
    \end{array} \right)    \; .
\end{align}
The transformed Hamiltonian in Eq.~(\ref{eq:XHX}) is obtained by using the relationship, $(\bsigma^{[4]} \cdot \bp)(\bsigma^{[4]} \cdot \bp)=\bp^2 1^{[4]}$ as
\begin{align}
    H_\text{KB}=
\begin{pmatrix}
    D_{1} & \frac{\bp^2_2}{2m_2}1\four & \frac{\bp^2_1}{2m_1}1\four & B_1 \\
    \frac{\bp^2_2}{2m_2}1\four & D_{2} & B_2 & \frac{\bp^2_1\bp^2_2}{8c^2m_1m_2^2}1\four\\
    \frac{\bp^2_1}{2m_1}1\four & B_3 & D_{3} & \frac{\bp^2_1\bp^2_2}{8c^2m_1^2m_2}1\four\\
    B_4 & \frac{\bp^2_2\bp^2_1}{8c^2m_1m_2^2}1\four & \frac{\bp^2_2\bp^2_1}{8c^2m_1^2m_2}1\four & D_{4}
\end{pmatrix} \; ,
    \label{eq:DCKB}
\end{align}
where the diagonal elements are
\begin{align}
\begin{split}
    D_{1}&=V1\four +U1\four \\
    D_{2}&=\frac{(\bsigma_2^{[4]}\bp_2)(V1\four +U1\four)(\bsigma_2^{[4]}\bp_2)}{4m_2^2c^2}-\frac{\bp_2^2}{2m_2} 1\four \\
    D_{3}&=\frac{(\bsigma_1^{[4]}\bp_1)(V1\four +U1\four)(\bsigma_1^{[4]}\bp_1)}{4m_1^2c^2}-\frac{\bp^2_1}{2m_1} 1\four \\
    D_{4}&=\frac{(\bsigma_1^{[4]}\bp_1)(\bsigma_2^{[4]}\bp_2)(V1\four +U1\four)(\bsigma_1^{[4]}\bp_1)(\bsigma_2^{[4]}\bp_2)}{16m_1^2m_2^2c^4}-\frac{m_{12}}{8m_1^2m_2^2c^2} \bp_1^2\bp_2^2 1\four \; .
\end{split}
    \label{eq:DCHdiag}
\end{align}
The anti-diagonal blocks, which carry the magnetic interactions, take the following form in
the transformed Hamiltonian
\begin{align}
\begin{split}
    B_1&=\frac{B(\bsigma_1^{[4]}\bp_1)(\bsigma_2^{[4]}\bp_2)}{4c^2m_1m_2} \\
    B_2&=\frac{(\bsigma_2^{[4]}\bp_2)B(\bsigma_1^{[4]}\bp_1)}{4c^2m_1m_2}\\
    B_3&=\frac{(\bsigma_1^{[4]}\bp_1)B(\bsigma_2^{[4]}\bp_2)}{4c^2m_1m_2}\\
    B_4&=\frac{(\bsigma_2^{[4]}\bp_2)(\bsigma_1^{[4]}\bp_1)B}{4c^2m_1m_2} \; .
    \label{eq:BreitKB}
\end{split}
\end{align}
The overlap matrix corresponding to the transformed Hamiltonian, Eq.~(\ref{eq:XHX}), is
\begin{align}
    S_\text{KB} 
    = 
    X^\dagger X 
    =
    \text{diag}\left(1\four,\frac{\bp_2^2}{4c^2m_2^2}1\four,\frac{\bp_1^2}{4c^2m_1^2}1\four,\frac{\bp_1^2\bp_2^2}{16c^4m_1^2m_2^2}1\four\right) \; .
    \label{eq:defSKB}
\end{align}

\subsection{Complex-coordinate rotation for positive-energy projection \label{ch:ccr}}
The complex-coordinate rotation transformation (CCR), used to define the positive energy projector (Sec.~III.B.1 of Paper~I),
scales the coordinates by a complex phase factor
\begin{align}
    x_i \ &\longrightarrow \ x_i\eem^{\iim \theta} \\
    \pd{}{x_i} \ &\longrightarrow \ \pd{}{x_i} \eem^{-\iim \theta} \\
    \frac{1}{r_{ij}} \ &\longrightarrow \  \frac{1}{r_{ij}} \eem^{-\iim \theta} \; ,
\end{align} 
and this transformation introduces only a simple complex scaling factor, $\eem^{-\iim\theta}$, 
\begin{align}
     B^\text{CCR}(\theta)=B\eem^{-\iim\theta} \; ,
\end{align}
in front of the Breit operator and also in front of the Breit matrix elements. 
In short, the Breit term is dilatation analytic under CCR. 
With this single modification in the calculation of the Breit matrix elements, the CCR positive-energy projector for the Dirac--Coulomb part of the problem is constructed and used as it is described in Paper~I (Sec.~III.B.1).

\subsection{Variational solution of the no-pair Dirac--Coulomb--Breit equation \label{ch:varDCB}}
We compute variational, no-pair Dirac--Coulomb (DC) and Dirac--Coulomb--Breit (DCB) energies and wave functions by direct diagonalization of the matrix representation of the  projected Hamiltonian, 
\begin{align}
    \bos{\Lambda}_+^\text{L} \bos{H}_\text{KB} \bos{\Lambda}_+^\text{R}  \bos{c}_i= E_i \bos{\Lambda}_+^\text{L} \bos{S}_\text{KB}\bos{\Lambda}_+^\text{R}  \bos{c}_i \; ,
\end{align}
where $\bos{H}_\text{KB}$ and $\bos{S}_\text{KB}$ denote the matrix representation of the corresponding operators, $\bos{c}_i$ and $E_i$ label the expansion coefficient vector and the energy eigenvalue, respectively.
It is important to note that the construction of the CCR projector (Secs. III.A--III.B of Paper I) assumes the solution of a non-Hermitian eigenvalue problem, so we build $\bos{\Lambda}_+^\text{L}$
and $\bos{\Lambda}_+^\text{R}$ from the left- and right-handed eigenvectors. For the cutting (or punching) projector  (Secs. III.B.2-3 of Paper I), the non-interacting problem is Hermitian, hence we may label $\bos{\Lambda}_+=\bos{\Lambda}^\text{R}_+$, and then $\bos{\Lambda}^\text{L}_+=\bos{\Lambda}_+^\dagger$. 

The Hamiltonian matrix is constructed either for the DC or the DCB operator and provides the positive-energy projected or no-pair DC or DCB energies, $E_\DC^\proj$ or $E_\DCB^\proj$, respectively. 
The non-linear parameters in Eq.~(\ref{eq:ECGbasis}) were optimized by minimizing the non-relativistic energy. This construction is expected to perform well for low-$Z$ systems. 
We have carried out test computations for further optimization of the non-linear parameters by minimization of the no-pair DCB energy (for the He, H$_2$, HeH$^+$, and H$_3^+$ systems studied in this paper), but no significant improvement was observed. 

Further test computations were carried out for the H$_2$ molecule regarding the contribution of non-relativistic basis functions corresponding to triplet spin states  that can mix with the singlet ground state functions ($LS$ coupling scheme), but the contribution of the triplet basis sector was negligible ($<  1$~n$\Eh$\ in the energy). Further details will be reported in future work. Regarding the helium atom, the non-relativistic ground state was computed with choosing $\bos{s}=0$ in the basis functions, Eq.~(\ref{eq:ECGbasis}). In test computations, this basis set was extended with $\bos{s}\neq 0$ functions, but their effect on the DC(B) energy was found to be very small.

All computations were carried out using double precision arithmetic (unless indicated otherwise) using the QUANTEN computer program. 
QUANTEN is an in-house developed program written using the Fortran90 programming language and contains several analytic ECG integrals, for recent applications have been reported in  Refs.~\cite{Ma18nonad,Ma18he2p,FeMa19EF,FeMa19HH,FeKoMa20,MaCa21,JeIrFeMa21,
IrJeMaMaRoPo21}. 
 
\subsection{Perturbative inclusion of the Breit interaction}
\noindent We have considered not only the variational but also the perturbative inclusion of the Breit interaction in the no-pair Dirac--Coulomb framework. The Rayleigh--Schrödinger perturbative corrections up to the first and second orders take the following form,
\begin{align}
    \langle B \rangle_n
    &=
    \left\langle %
      \bos{\Psi}^{\text{proj}}_{\text{DC},n} \big|  X^\dagger B(1,2) X \big| \bos{\Psi}^{\text{proj}}_{\text{DC},n}
    \right\rangle  
    \label{eq:pt1}
    \\
    \mathcal{P}^{(2)}_n\lbrace B\rbrace
    &=
    \langle B \rangle_n +
    \sum_{i \neq n} 
    \frac{%
      \abs{%
        \left\langle%
          \bos{\Psi}^{\text{proj}}_{\text{DC},i} 
          \big| X^\dagger B(1,2) X \big| 
          \bos{\Psi}^{\text{proj}}_{\text{DC},n}
        \right\rangle
      }^2
    }{%
      E^{\text{proj}}_{\text{DC},i}-E^{\text{proj}}_{\text{DC},n}
    }  
    \label{eq:pt2}
    \; ,
\end{align}
where $\bos{\Psi}^{(\text{proj})}_{\text{DC},n}$ and $E^{(\text{proj})}_{\text{DC},n}$ is the $n$th no-pair DC wave function and energy. 
Since we calculate perturbative corrections for the no-pair DC problem, Brown--Ravenhall or negative-energy states do not enter the expressions.

Equations~(\ref{eq:pt1}) and (\ref{eq:pt2}) are valid if a Hermitian positive-energy projector (cutting, punching or determinant projector of Paper~I) is used to set up the no-pair DC matrix. During the present work, the reported perturbative results were computed with the cutting projector. 
The implementation was generalized also for the CCR projector and the tested numerical results agree to all reported digits with the cutting projector values.

The first and second-order perturbative corrections were evaluated for comparison with the no-pair DCB energy, and they (and their good numerical agreement with the no-pair DCB energy) are shown in the Supplementary Material.

\clearpage
\section{Numerical results and discussion}
\subsection{Discussion of the projection techniques \label{ch:proj}}
Currently, the most rigorous positive-energy projection approach for methods using explicitly correlated basis functions is based on the complex-coordinate rotation (CCR) technique proposed by Bylicki, Pestka, and Karwowski \cite{ByPeKa08} for the solution of the Dirac--Coulomb problem of atoms. In Paper~I~\cite{JeFeMa21dc}, we have generalized this CCR projection approach to molecular computations with the DC Hamiltonian. 
We adapt the molecular approach to the DCB problem in the present work, and report observations in the following paragraphs.
The DC(B) Hamiltonian is dilatation analytic for atoms, but not for molecules with fixed nuclei. 

We have considered two approaches for molecular computations. The first approach is a non-dilatation analytic (`nda') approach, which directly accounts for the non-dilatation analytic property of the Hamiltonian due to the Coulomb interaction with the external charges (nuclei). 

A second possible approach for molecules relies on a na\"ive dilatation use of the CCR technique as if the Hamiltonian was dilatation analytic (`da'), \emph{i.e.,} the positions of the fixed nuclei are complex scaled together with the active electronic degrees of freedom. According to Moiseyev \cite{moiseyevNonHermitianQuantumMechanics2011}, this dilatation analytic approach is also appropriate, but for accurate results a perturbative correction for the `back rotation' of the nuclear positions is necessary. In Paper~I \cite{JeFeMa21dc}, this perturbative correction was found to be small for the imaginary part of the DC energy (that is 0 for a bound state) and negligibly small for the real part of the DC energy for sufficiently small CCR angles. Any $\theta$ CCR angle is appropriate for constructing the projector that is sufficiently large for a clear identification of the positive-energy branch of the non-interacting energies (Fig.~1 of Paper~I), but not too large for the finite basis set, \emph{i.e.,} the finite basis set error, which increases with $\theta$, remains small.
The advantage of the `da-CCR' approach is that it does not require the evaluation of complex-valued Coulomb integrals that contain the complex incomplete gamma function, which can be evaluated only with 12-digit precision in our current implementation. 

For low-$Z$ systems, studied in the present work, we have found a simple energy cutting projection technique (`cutting') also appropriate for a (sub-)parts-per-billion (ppb) computation of the DC energy. The cutting approach is technically and numerically the simplest one, and it was always performed as a first test computation. 

An overview and comparison of the numerical performance of the projection techniques for the no-pair DCB energy is provided in Tables~\ref{tab:heCCR} and \ref{tab:h2CCR} for the example of the helium atom and the hydrogen molecule.

Table~\ref{tab:heCCR} presents
the angle dependence of the CCR-projected DCB energies for the ground state of the helium atom and shows also the result of the simple `cutting' projector. For all $\theta\in[10^{-8},10^{-2}]$, the positive-energy non-interacting states could be separated, and the finite basis error for this interval was sufficiently small, hence the real parts of the DCB energy show a difference less than 1 in $10^9$ (sub-ppb).
The double precision arithmetic is sufficient for achieving ppb precision, but we have performed the computation using quadruple precision to be able to see the details of the numerical behaviour of the different projection techniques.

Table~\ref{tab:h2CCR} shows the numerical behaviour of the different projection techniques for the example of the ground state of the H$_2$ molecule. We observe a similar behaviour for the nda-CCR and cutting projectors as for the da-CCR and cutting projectors for the helium atom (Table~\ref{tab:heCCR}), respectively. Regarding the da-CCR projector, we used it without the perturbative correction for the back rotation of the nuclei, similarly to the DC problem in Paper I \cite{JeFeMa21dc}. For sufficiently small CCR angles, the real part of the da-CCR energy is in an excellent numerical agreement with the nda-CCR energy, since the real part for the perturbative back rotation scales with $\sim\theta^2$ (Paper~I). At the same time, we see a substantial difference in the imaginary part of the da-CCR and nda-CCR energies. This difference can be understood by noticing that the perturbative back rotation for the imaginary part scales with $\sim\theta$.

All in all, the tested positive-energy projection techniques provide the (real part of the) bound-state energy for the example systems (He and H$_2$) with a sub-ppb difference.

\begin{table}[h]
  \caption{%
    Testing the positive-energy projection techniques for the Dirac--Coulomb--Breit energy of the ground state of the helium atom. 
    The CCR angle dependence of the dilatation-analytic energy, $E^\mathrm{da}_\mathrm{DCB}$, 
    is shown in comparison with the energy cutting approach, $E^\mathrm{cutting}_\mathrm{DCB}$.
    Quadruple precision arithmetic and $N_\text{b}=300$ ECG functions were used.
    All energies are in $\Eh$ units.
    \label{tab:heCCR}}
 \begin{tabular}{@{}l lc r @{}}
    \hline\hline\\[-0.35cm]
    & \multicolumn{3}{c}{proj = da($\theta$)} \\
    \cline{2-4} \\[-0.3cm]
    \multicolumn{1}{c}{$\theta$} &	
    \multicolumn{1}{c}{$\mathrm{Re} (E^\mathrm{proj}_\DCB)$}  & \hspace{0.2cm} &
    \multicolumn{1}{c}{$\mathrm{Im} (E^\mathrm{proj}_\DCB)$} \\
    \cline{1-4}\\[-0.3cm]
     0.000 000 01	 & 	$-$2.903 828 970 048 & &  7.35 $\cdot 10^{-15}$ \\
     0.000 000 1	 & 	$-$2.903 828 970 048 & &  7.35 $\cdot 10^{-14}$ \\
     0.000 001	 & 	$-$2.903 828 970 048 & &  7.35 $\cdot 10^{-13}$ \\
     0.000 01	 & 	$-$2.903 828 970 048 & &  7.35 $\cdot 10^{-12}$ \\
     0.000 1	     & 	$-$2.903 828 970 048 & &  7.35 $\cdot 10^{-11}$ \\
     0.001	     & 	$-$2.903 828 970 048 & &  7.35 $\cdot 10^{-10}$ \\
     0.01	     & 	$-$2.903 828 970 071 & &  7.35 $\cdot 10^{-9}$\hspace{0.15cm} \\
     0.1	     & 	$-$2.903 828 971 461 & &  7.47 $\cdot 10^{-8}$\hspace{0.15cm} \\
     0.2	     & 	$-$2.903 828 969 543 & &  1.50 $\cdot 10^{-7}$\hspace{0.15cm} \\
     0.5	     & 	$-$2.903 829 132 724 & &  2.56 $\cdot 10^{-8}$\hspace{0.15cm} \\
    \cline{1-4}\\[-0.3cm]
    proj = cutting:  & & & \\
    $E^\mathrm{proj}_\DCB$  &  $-$2.903 828 970 048	&	&  \multicolumn{1}{c}{0} \\
    \hline\hline
  \end{tabular}
\end{table}

\begin{table}
  \caption{%
    Testing the positive-energy projection techniques for the  Dirac--Coulomb--Breit energy of the ground-state of the H$_2$ molecule with fixed protons ($R_\text{pp}=1.4$~bohr).
    The dilatation analytic ($E^\mathrm{da}_\DCB$), the non-dilatation analytic ($E^\mathrm{nda}_\DCB$), and the energy cutting ($E^\mathrm{cutting}_\DCB$) techniques are compared. 
    Quadruple precision arithmetic and $N_\text{b}=700$ ECG functions were used.
    All energies are in $\Eh$ units.
  }
  \label{tab:h2CCR}
    \begin{tabular}{@{}llcc@{}}
    \hline\hline\\[-0.35cm]
    \multicolumn{1}{c}{$\theta$} & 
    \multicolumn{1}{c}{$\mathrm{Re} (E^\mathrm{proj}_\DCB)$}  & 
     & 
    \multicolumn{1}{c}{$\mathrm{Im} (E^\mathrm{proj}_\DCB)$} \\  
    \cline{2-4} \\[-0.3cm]
    & \multicolumn{3}{c}{proj = nda($\theta$)} \\
    \cline{1-4}\\[-0.3cm]
    0.000 000 1	 & $-$1.174 486 710 978 & &	5.55 $\cdot 10^{-14}$ \\
    0.000 001	 &  $-$1.174 486 710 978 & & 5.55 $\cdot 10^{-13}$ \\
    0.000 01	 & 	$-$1.174 486 710 982  & &  5.55 $\cdot 10^{-12}$ \\
    0.000 1	     & $-$1.174 486 711 397	 & &	5.54 $\cdot 10^{-11}$ \\
    0.001	     &  $-$1.174 486 753 444	 & &	4.63 $\cdot 10^{-10}$ \\
     \cline{1-4}\\[-0.3cm]
     & \multicolumn{3}{c}{proj = da($\theta$)} \\
    \cline{2-4}  \\[-0.3cm]
    0.000 000 1	 & $-$1.174 486 710 978 & & 	7.14 $\cdot 10^{-8}$ \\
    0.000 001	 & $-$1.174 486 710 978 & &  7.14 $\cdot 10^{-7}$ \\
    0.000 01	 & $-$1.174 486 710 979 & & 	7.14 $\cdot 10^{-6}$   \\
    0.000 1	     & $-$1.174 486 711 043  & & 7.14 $\cdot 10^{-5}$ \\
    0.001        & $-$1.174 486 717 484 & &  7.14 $\cdot 10^{-4}$ \\
    \cline{1-4}\\[-0.3cm]
    proj = cutting:  & & & \\
    $E^\mathrm{proj}_\DCB$  &  $-$1.174 486 710 978 	&	&  \multicolumn{1}{c}{0}  \\
    \hline\hline
    \end{tabular}
\end{table}

\clearpage
\subsection{Discussion of the numerical results}
The energies computed with the inclusion of the Breit correction, Eq.~(\ref{eq:breit}),
are summarized in Table~\ref{tab:compare}.  
The table collects the results obtained with the largest basis sets 
for the helium atom, and for the ground electronic state of the H$_2$, HeH$^+$ and H$_3^+$ molecular systems near their equilibrium geometry. 

Regarding the convergence details, we consider the no-pair DC plus first-order perturbative Breit energy, $E_\DC^\proj+\langle B\rangle_\DC$, converged better than 5-10~n$\Eh$ for all systems studied. 
The no-pair DCB energy, $E_\DCB^\proj$, converged much slower, our largest basis results are probably converged only within 10~n$\Eh$\ for H$_2$, H$_3^+$ and He $2\ ^1S_0$, but only on the order of ca.~100~n$\Eh$\ for He $1\ ^1S_0$ and HeH$^+$.
In Table~\ref{tab:compare}, one more digit is shown beyond the converged values and 
detailed convergence tables are provided in the Supplementary Material. 
The parameterization for all basis sets used in these computations were obtained by minimization of the non-relativistic energy.

\begin{table}[h]
\caption{%
    The no-pair DC energy with first-order perturbative Breit correction, 
    $E_\DC^\proj +\langle B \rangle_\DC$ in $\Eh$, and
    the no-pair DCB energy, $E_\DCB^\proj$ in $\Eh$.
    The differences, $\delta$ and $\delta'$ in n$\Eh$, with respect to 
    the non-relativistic energy with the leading-order ($\alpha^2$) perturbative relativistic energy, $E_\DCB^{(2)}$, Eq.~(\ref{eq:ptalpha2}), is also shown. The estimated convergence error appears in the last digit.
    \label{tab:compare}
}
  \begin{tabular}{@{}l lr lr lr@{}}
    \hline\hline\\[-0.35cm]
    \multicolumn{1}{c}{} &
     & &
    \multicolumn{1}{c}{$E_\DC^\proj+\langle B \rangle_\DC$} &
    \multicolumn{1}{c}{$\lbrace\delta\rbrace$ $^\mathrm{a}$} &
    \multicolumn{1}{c}{$E_\DCB^\proj$}&
    \multicolumn{1}{c}{$\lbrace\delta'\rbrace$ $^\mathrm{b}$}\\
    \hline\\[-0.35cm]
    H$_2$ $^\mathrm{c}$ 
    &&
    & {$-$1.174 486 665} & \{2\} 
    & {$-$1.174 486 721} & \{$-$54\}  \\
    H$_3^+$ $^\mathrm{c}$
    &&  
    & $-$1.343 847 416  & \{0\}  
    & $-$1.343 847 498  & \{$-$82\} \\
    HeH$^+$ $^\mathrm{c}$
    & &  
    & $-$2.978 808 200  & \{$-$20\}  
    & $-$2.978 808 77  & \{$-$590\} \\    
    He ($1\ ^1S_0$)       &&                
    & $-$2.903 828 333 & \{$-$22\}
    & $-$2.903 829 02 & \{$-$710\} \\
    He ($2\ ^1S_0$)                 
    && 
    & $-$2.146 082 379   & \{$-$11\}  
    & $-$2.146 082 424   & \{$-$56\} \\
    \hline\hline
  \end{tabular}
  \begin{flushleft}
    $^\mathrm{a}$  
      $\delta=E_\DC^\proj+\langle B\rangle_\DC-E^{(2)}_\DCB$. \\
    $^\mathrm{b}$  
      $\delta'=E_\DCB^\proj-E^{(2)}_\DCB$. \\
    $^\mathrm{c}$ 
      Electronic ground state for nuclear-nuclear distances
      $R_\text{eq}=1.4$~bohr, 1.65~bohr, and 1.46~bohr for H$_2$, H$_3^+$, and HeH$^+$, respectively.
  \end{flushleft}
\end{table}

For small and light systems, the most accurate results (and results in good agreement with experiment) have been reported using perturbative techniques. For this reason, Table~\ref{tab:compare} also shows the deviation ($\delta$ and $\delta'$) of the computed energies and the leading-order perturbative relativistic energy, \emph{i.e.,} the sum of the non-relativistic energy, $E_\text{nr}$, and the expectation value of the Breit--Pauli Hamiltonian (the $\alpha^2$ term) with the non-relativistic wave function, $\Psi_\text{nr}$ (see for example, Ref.~\cite{DyFaBook07}),
\begin{align}
  E_\DCB^{(2)}
  =
  E_\text{nr}
  +
  \alpha^2
  \langle
    \Psi_\text{nr} |
      H_\DC^{(2)}+H_\text{B}^{(2)}
    |\Psi_\text{nr}
  \rangle
  \label{eq:ptalpha2}
\end{align}
with
\begin{align}
  H_\DC^{(2)}
  &=
  -\frac{1}{8}\sum_{i=1}^N (\bos{\nabla}_i^2)^2
  + 
  \frac{\pi}{2}\sum_{i=1}^N\sum_{A=1}^{N_\mathrm{nuc}} Z_A \delta(\br_{iA}) 
  -\pi \sum_{i=1}^N\sum_{j>i}^N \delta(\br_{ij}) 
  \label{eq:fwptdc}
  \\
  H_\text{B}^{(2)}
  &=
  H_\text{OO} + 2\pi \sum_{i=1}^N\sum_{j>i}^N \delta(\br_{ij})
  \label{eq:fwptbreit}
\end{align}
and
\begin{align}
  H_\text{OO}
  =
  -\sum_{i=1}^N\sum_{j>i}^N 
  \frac{1}{2r_{ij}}
  \left(%
    \bos{p}_i\bos{p}_j
    +
    \frac{\bos{r}_{ij}(\bos{r}_{ij}\bos{p}_i)\bos{p}_j}{r_{ij}^2}
  \right)  \; .
  \label{eq:fwptoo}
\end{align}
We note that the speed of light in Secs.~\ref{ch:intro} and \ref{ch:npDCB} can be written as $c=\alpha^{-1} a_0\Eh/\hbar$ in Hartree atomic units. In the perturbative expressions, Eqs.~(\ref{eq:ptalpha2})--(\ref{eq:fwptoo}), the $H_\DC^{(2)}$, $H_\text{B}^{(2)}$, and $H_\text{OO}$ terms are understood in this context to have units of $\Eh$ (hartree).

We may observe in Table~\ref{tab:compare} that the no-pair DCB energy has a very large deviation ($\delta'$), whereas
the $E_\DC^\proj+\langle B\rangle_\DC$ energy has a smaller, but non-negligible, difference ($\delta$) from the $E^{(2)}_\DCB$ leading-order relativistic energy.
The sign and the order of magnitude of the deviation is not immediately obvious and motivates further analysis. 
In Table~III of Paper~I, we observed that it is necessary to go beyond the leading-order ($\alpha^2$) perturbative correction and include also the $\alpha^3$-order non-radiative QED term for the two-Coulomb-photon exchange to have a good agreement with the no-pair Dirac--Coulomb energy. 
For a better understanding of the Breit results, let us consider the effect of the Breit correction, instead of the full energy.

\begin{table}[h]
  \caption{%
     Comparison of the Breit correction, in $\mu\Eh$, obtained from different computations. 
     The estimated convergence error appears in the last digit.
     All values correspond to $\alpha=\alpha_0=137.035 999 084$~\cite{codata18}.
     \label{tab:breitcompare}
  }
  \begin{tabular}{@{}l@{\ \ \ } l@{\ \ \ }l@{\ \ \ }l@{\ \ \ }l@{\ \ \ } l@{\ \ \ }l@{\ \ \ }l@{}}
    \hline\hline\\[-0.35cm]
	  &	H$^-$	&	He	&	Li$^+$	&	Be$^{2+}$	&	H$_2$	&	H$_3^+$	&	HeH$^+$	\\
    \hline\\[-0.35cm]	  
    $E_\DCB^\proj-E_\DC^\proj$
      &	0.436	&	27.61	&	150.67	&	449.3	&	3.033	&	3.029	&	25.95	\\
    $\langle B\rangle_\text{DC}$ $^\text{a}$	
      &	0.448	&	28.298	&	156.354	&	464.60	&	3.089	&	3.111	&	26.436	\\
    $\langle B\rangle_\text{P}$ $^\text{a,b}$	
      &	0.445	&	28.209	&	155.907	&	463.11	&	3.074	&	3.095	&	26.347	\\
    $\alpha^2\langle H^{(2)}_\text{B}\rangle_\text{nr}$ $^\text{a,b}$  \cite{Dr06,PuKoPa17,JeIrFeMa21,JeFeMa21}
      &	0.4435	&	28.1749	&	155.7861	&	462.7471	&	3.0655	&	3.0877	&	26.313	\\
    $\langle B_\omega\rangle_\text{MCDF}$ $^\text{a,c}$ \cite{PaGr90}
      &	0.4	&	28	&	(n.a.)	&	460   &	(n.a.)	&	(n.a.)	&	(n.a.)	\\
    \hline\hline\\[-0.4cm]
  \end{tabular}
  \begin{flushleft}
   $^\text{a}$~%
   $\langle O\rangle_X$: expectation value of the $O$ operator with the $X$: projected Dirac--Coulomb (DC), Pauli (P), non-relativistic (nr), and multi-configuration Dirac--Fock (MCDF) wave function. \\
  $^\text{b}$~%
    $\langle B\rangle_\text{P}$ equals (by mathematically equivalent rearrangements) $\alpha^2\langle H^{(2)}_\text{B}\rangle_\text{nr}$, so the same result is obtained, 
    as if $\langle H^{(2)}_\text{B}\rangle_\text{nr}$ was `directly' evaluated in the same basis set as $\langle B\rangle_\text{P}$. 
    The reference values for $\langle H^{(2)}_\text{B}\rangle_\text{nr}$
    were obtained by techniques that account for the cusp of the non-relativistic (nr) wave function and enhance the convergence of the singular operator in $H^{(2)}_\text{B}$, Eq.~(\ref{eq:fwptbreit}). \\
  $^\text{c}$~%
  Perturbative corrections including the frequency-dependence of the Breit operator taken from  Ref.~\cite{PaGr90}. Eq.~(\ref{eq:onephx}) is the relevant equation in this work that contains the sum of the Coulomb and the $B_\omega$ frequency-dependent Breit interactions.
  \end{flushleft}
\end{table}

Table~\ref{tab:breitcompare} shows Breit correction values obtained from different computations. (For the sake of further analysis, we have also included the ground state of Li$^+$ and Be$^{2+}$ in the table.)
First of all, we observe that the bulk of the Breit contribution (first 1-2 significant digits) agree in all computations.

A good agreement is observed even with the $\langle B_\omega\rangle_\text{MCDF}$ multi-configuration Dirac--Hartree--Fock value \cite{PaGr90}, in which $B_\omega$ includes not only the non-retarded Breit term (as in our work), but accounts also for the frequency dependence of the interaction. Further digits are not available from that computation, but the available information provides us an independent check, assuming that the frequency dependence (and the different projector) has only a small effect for the systems studied.
Apart from this set of values, all other values listed in the table correspond to the unretarded Breit interaction,  Eq.~(\ref{eq:Breit2}).
Regarding the uncertainty of the digits shown in the table,
the estimated convergence error appears in the last digit given for $E_\DCB^\proj-E_\DC^\proj$ and $\langle B\rangle_\text{DC}$. 
All digits shown for the reference value of $\alpha^2\langle H^{(2)}_\text{B}\rangle_\text{nr}$ are significant, \emph{i.e.,} converged for that quantity. 

For $\langle B\rangle_\text{P}=\langle \Psi_\text{P}| B \Psi_\text{P}\rangle$, we show the relevant number of digits of the numerical value obtained with the Pauli wave function constructed from the non-relativistic wave function optimized in this work. 
The Pauli wave function \cite{sucherPhD1958} is an approximation to the no-pair DC wave function, it can be constructed by using the Eq.~(\ref{eq:XHX}) kinetic balance condition and by filling up the appropriate spin-components with the (normalized) non-relativistic wave function in  all (ll, ls, sl, and ss) blocks. So, we have evaluated $\langle B\rangle_\text{P}$ by using the non-relativistic wave function in this way in our sixteen-component Breit implementation. 
Furthermore, it can be shown that $\langle \Psi_\text{P}| B | \Psi_\text{P}\rangle$ can be exactly rearranged to $\alpha^2\langle H_\text{B}^{(2)}\rangle_\text{nr}=\alpha^2\langle \Psi_\text{nr}| H_\text{B}^{(2)} | \Psi_\text{nr} \rangle$ (normalization of the Pauli wave function gives contribution only at higher $\alpha$ orders) \cite{sucherPhD1958}.
In our implementation, the $\langle B\rangle_\text{P}=\alpha^2\langle H_\text{B}^{(2)}\rangle_\text{nr}$  mathematical identity is fulfilled to machine precision, if the same non-relativistic wave function is used in the two computations. 
Although the same underlying ECG integral routines are used for the evaluation of the two expressions, the two different computations assume the combination of different indices, etc. (compare Sec.~\ref{ch:npDCB} in the one hand and Eqs.~(\ref{eq:fwptbreit})--(\ref{eq:fwptoo}) on the other), so this was an important check of our Breit implementation.

The $\langle B\rangle_\text{P}$ values reported in the table correspond to the ECG basis that was obtained by minimization of the non-relativistic energy and also used as a `spatial' basis in the no-pair DC and DCB computations. As it was pointed out, these $\langle B\rangle_\text{P}$ values agree to machine precision with the direct evaluation of $\alpha^2\langle H_\text{B}^{(2)}\rangle_\text{nr}$ in the same basis (`own basis'). 
At the same time, these values are not converged with respect to the basis set size due to the slowly convergent expectation value of the singular operator, $\delta(\bos{r}_{12})$, which can be most easily identified in Eq.~(\ref{eq:fwptbreit}), in a Gaussian basis not satisfying the cusp condition \cite{JeIrFeMa21}.
There are special `regularization' techniques \cite{Dr81,PaCeKo05,JeIrFeMa21} that can be used to improve the convergence of the expectation value for the singular operator, $\delta(\bos{r}_{ij})$ (as well as for other singular operators in $H_\DC^{(2)}$ in Eq.~(\ref{eq:fwptdc})). 

The well-converged perturbative values are compiled from literature values in the $\alpha^2 \langle H_\text{B}^{(2)}\rangle_\text{nr}$ line of Table~\ref{tab:breitcompare}. We also note that the $\alpha^2\langle H_\text{B}^{(2)} \rangle_\text{nr}$ values for HeH$^+$ and  H$_3^+$ were taken from Refs.~\cite{JeFeMa21}  and \cite{JeIrFeMa21}, respectively, and were computed with ECG functions and the so-called integral transformation technique that enhance the convergence by accounting for the missing cusp effects \cite{PaCeKo05,JeIrFeMa21}.
Further numerical details and quantities used to calculate the perturbative corrections are collected in the Supplementary Material.


After this introduction, we are in a position to compare the $\alpha^2\langle H_\text{B}^{(2)}\rangle_\text{nr}$ value, which is a quadratic function in $\alpha$ ($\langle H_\text{B}^{(2)}\rangle_\text{nr}$ in Hartree atomic units, which we use in this work, is independent of $\alpha$), with the no-pair, sixteen-component results, $E_\DCB^\proj-E_\DC^\proj$ and $\langle B\rangle_\text{DC}$ that contain non-radiative QED contributions, \emph{i.e.,} contributions beyond $\alpha^2$. To quantify this dependence, we solved the no-pair DC(B) equation for several $\alpha$ values in the interval $\alpha\in[0.7,1.6]\alpha_0$ with $\alpha_0=1/137.$035~999~084 \cite{codata18}.
For much smaller $\alpha$ values, the corrections are too small for our numerical precision, for much larger $\alpha$ values, the spatial basis set taken from a non-relativistic optimization procedure is insufficient.

\begin{figure}
    \centering
    \includegraphics[scale=0.7]{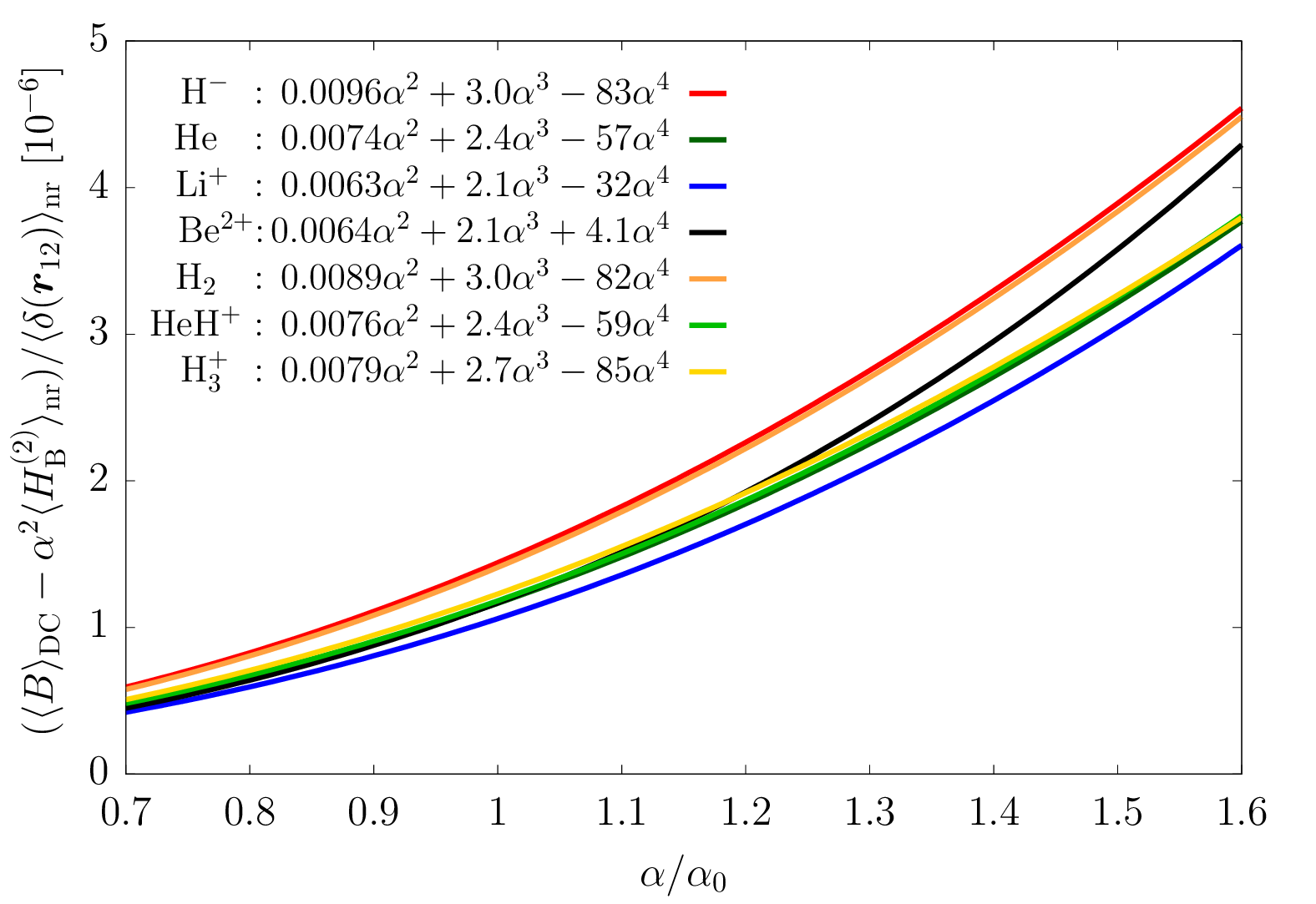}
    \caption{%
      Dependence of the Breit correction to the no-pair DC energy, $\langle B\rangle_\text{DC}$, on the value of the $\alpha$ coupling constant of the electromagnetic interaction. 
      Hartree atomic units are used and 
      $\alpha_0$ labels $1/137.$035~999~084 \cite{codata18}. 
      The data points, used for fitting the polynomials, were computed at the $\alpha=1/(\alpha_0+n)$, $n=-50, \dots, 50$ values.      
      The $\langle H_\text{B}^{(2)} \rangle_\mathrm{nr}$ and the $\langle \delta(\bos{r}_{12})\rangle_\text{nr}$ values compiled from Refs.~\cite{Dr06,PuKoPa17,JeFeMa21,JeIrFeMa21} are listed in Table~\ref{tab:breitcompare} and in the Supplementary Material. 
      \label{fig:BreitPTalphafit}
    }
\end{figure}

Figure~\ref{fig:BreitPTalphafit} shows the $\langle B\rangle_\text{DC}(\alpha)-\alpha^2\langle H_\text{B}^{(2)}\rangle_\text{nr}$ difference and the $b_2\alpha^2+b_3\alpha^3+b_4\alpha^4$ polynomial fitted to the difference. 
In each fit, there is a small $b_2\alpha^2$ `relativistic offset' that can be attributed to imperfections in the convergence of the $\langle B\rangle_\text{DC}$ data. 
The data is normalized with $\langle \delta(\bos{r}_{12})\rangle_\text{nr}$ that brings the values of all helium-like ions ($Z=1,2,3,4$) and the three molecules (H$_2$, HeH$^+$, and H$_3^+$) to the same scale in the figure. The $b_3$ coefficients are similar in the different systems, but there is some system dependence that requires further study and in the first place, better converged $\langle B\rangle_\text{DC}$ data. All in all, we may conclude that the leading-order non-radiative QED contribution due to an (unretarded) Breit photon is approximately described by
\begin{align}
  \langle B\rangle_\text{DC}^{(3)}
  \approx
  2.5 \alpha^3 \langle \delta(\bos{r}_{12})\rangle_\text{nr} \; .
\end{align}
The $\langle \delta(\bos{r}_{12})\rangle_\text{nr}$ proportionality is in agreement with Sucher's result (Chapter~V of Ref.~\cite{sucherPhD1958}), but Sucher's coefficient for the positive-energy contribution (for singlet states) is different, Eq.~(5.64) of Ref.~\cite{sucherPhD1958}:
\begin{align}
  \alpha^3\varepsilon_\text{CB,unret}^{++}
  =4\left(\frac{\pi}{2}+1\right) \alpha^3 \langle \delta(\bos{r}_{12}) \rangle_\text{nr} 
  \approx 
  10.3 \alpha^3 \langle \delta(\bos{r}_{12}) \rangle_\text{nr} 
  \; . 
  \label{eq:cbpp}
\end{align}
We do not have a conclusive explanation for this discrepancy. 
It is interesting to note however that the single- and double-pair contributions, Eqs.~(5.21a)--(5.21b) of Ref.~\cite{sucherPhD1958}, to the (unretarded) expression in Eq.~(\ref{eq:cbpp}) (for singlet states) sum to
\begin{align}
  \alpha^3\varepsilon_\text{CB,unret}
  =
  \left(%
    4 -2\text{ln}2
    \right) 
  \alpha^3 \langle \delta(\bos{r}_{12}) \rangle_\text{nr} 
  \approx 
  3.40 \alpha^3 \langle \delta(\bos{r}_{12}) \rangle_\text{nr} \; ,
\end{align}
and observe that Sucher approximates the intermediate states with free-electron states, while they are one-electron states in the field of the nuclei (without electron-electron interactions) in our computations. Further work, including the pair corrections in our computations will help to clarify these aspects.

The $\alpha^4$-order contribution is not expected to be well approximated with a constant multiple of $\langle\delta(\bos{r}_{12})\rangle_\text{nr}$. Nevertheless, the $b_4\alpha^4$ term is included in the fit (Fig.~\ref{fig:BreitPTalphafit}), and we can indeed observe a significant variation of $b_4$ over the different systems studied.

Based on these observation, we can address the good numerical agreement of $E_\DC^\proj+\langle B\rangle_\DC$ and
$E^{(2)}_\text{DCB}$ (Table~\ref{tab:compare}).
The leading-order non-radiative QED correction in $E_\DC^\proj$ (Fig.~3 in Paper~I) and in $\langle B\rangle_\DC$ have an opposite sign and a similar order of magnitude. The sum of their contribution can be described to leading order in $\alpha$ as $(b_3+c_3)\alpha^3 \approx -\alpha^3 \langle\delta(\bos{r}_{12})\rangle$ that corresponds 
for the physical $\alpha=\alpha_0$ value 
to values on the order of $(-1)$--$(-7)$~n$\Eh$ for the hydrogenic systems and $-40$~n$\Eh$ for the $Z=2$ systems. These values, together with the convergence estimates---5 and 10~n$\Eh$ for $Z=1$ and $Z=2$ systems, respectively---, and the observed remaining higher-order contributions in $E_\DC^\proj$ (Sec. IV.C of Paper~I) provides some insight to the numerical agreement of $E_\DC^\proj+\langle B\rangle_\DC$ and
$E^{(2)}_\text{DCB}$ in Table~\ref{tab:compare}.

The $\alpha$ dependence of the no-pair Dirac--Coulomb--Breit energy is much less understood.
First of all, it is important to note that the `bulk of the correction' is correct and it is in agreement with other Breit correction results (Table~\ref{tab:breitcompare}). At the same time, we observe a slow convergence of $E_\DCB^\proj$ with the basis-set size. Furthermore, $E_\DCB^\proj-E_\DC^\proj$ is `much' smaller than $\langle B\rangle_\DC$, and the absolute value of the deviation is `large' in comparison with the typical deviations of the perturbative theory and experiment. By adding more ECG functions (that is currently limited by the double precision arithmetic that we use), we may expect some further decrease of the $E_\DCB^\proj$ energy, \emph{i.e.,} increase of the absolute value of the deviation, due to the (near-)variational property of the no-pair computations and the already good convergence of the $E_\DC^\proj$ energy. 

\begin{figure}
    \centering
    \includegraphics[scale=0.7]{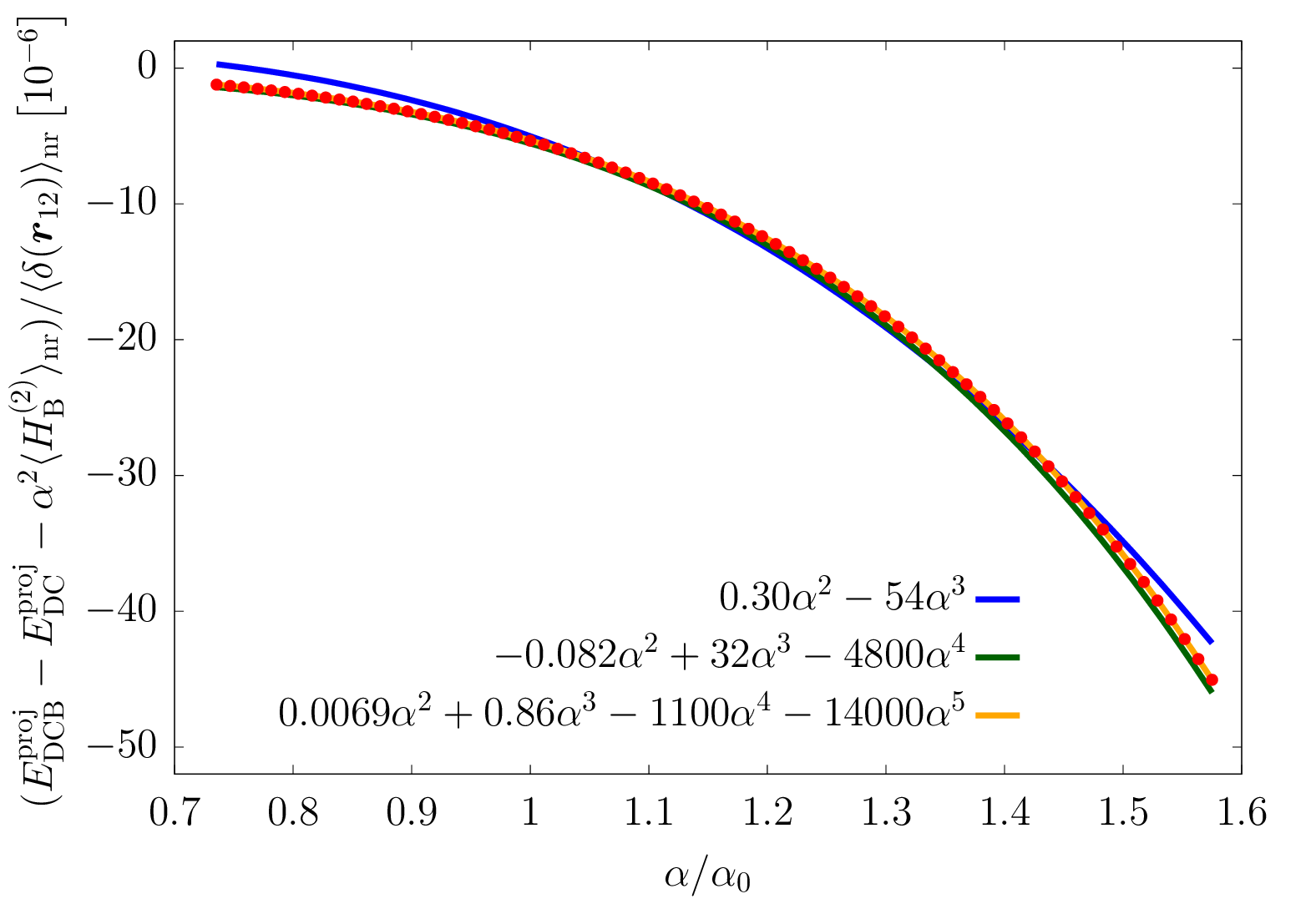} \\
    \caption{%
      Dependence of the no-pair Dirac--Coulomb--Breit (DCB) energy on the value of the $\alpha$ coupling constant of the electromagnetic interaction for the example of the ground state of the helium atom. 
      Hartree atomic units are used and 
      $\alpha_0$ labels $1/137.$035~999~084 \cite{codata18}. 
      The $\alpha$ dependence of the no-pair Dirac--Coulomb energy, $E_\DC^\proj$, is shown in Fig.~3 of Paper~I.
      The $\langle H_\text{B}^{(2)} \rangle_\mathrm{nr}$ and $\langle \delta(\bos{r}_{12})\rangle_\text{nr}$ values compiled from Ref.~\cite{Dr06} are listed in Table~\ref{tab:breitcompare} and in the Supplementary Material. 
      %
      \label{fig:DCBalphafit}
    }
\end{figure}

It is interesting to note that the second-order Rayleigh--Schrödinger perturbative Breit correction, Eqs.~(\ref{eq:pt1}) and (\ref{eq:pt2}), to the no-pair DC energy practically recovers the no-pair DCB result (the numerical values are provided in the Supplementary Material). This suggests that the deviation could be understood in terms of a perturbative correction (from the positive energy space) due to the exchange of two (unretarded) Breit photons. Sucher estimated this value to be (for singlet states) $-\frac{\pi}{2} \alpha^3\langle \delta(\bos{r}_{ij})\rangle_\text{nr}$ in Eq.~(6.9b++) of Ref.~\cite{sucherPhD1958}. The sign of the deviation is in agreement with our numerical result, but it is an order of magnitude(!) smaller, than the difference found in our computations. The comparison is further complicated by the fact that Sucher approximates the intermediate states with free-electron states. By collecting the pair corrections corresponding to Sucher's free-electron intermediate-state approximation \cite{sucherPhD1958}, we may observe the so-called Araki--Sucher term and an $\text{ln}\alpha$ dependence.
For the example of the helium atom, we have studied the $\alpha$ dependence of the no-pair DCB and DC energies (Fig.~\ref{fig:DCBalphafit}). Surprisingly, the fitting coefficients in  $\sum_{n=2}^m t'_n \alpha^n$ change significantly by increasing the maximal polynomial order from $m=3$ to 4. Inclusion of an $\alpha^3 \text{ln} \alpha$ term did not improve the quality of the fit.

\begin{figure}
    \centering
    \includegraphics[scale=0.7]{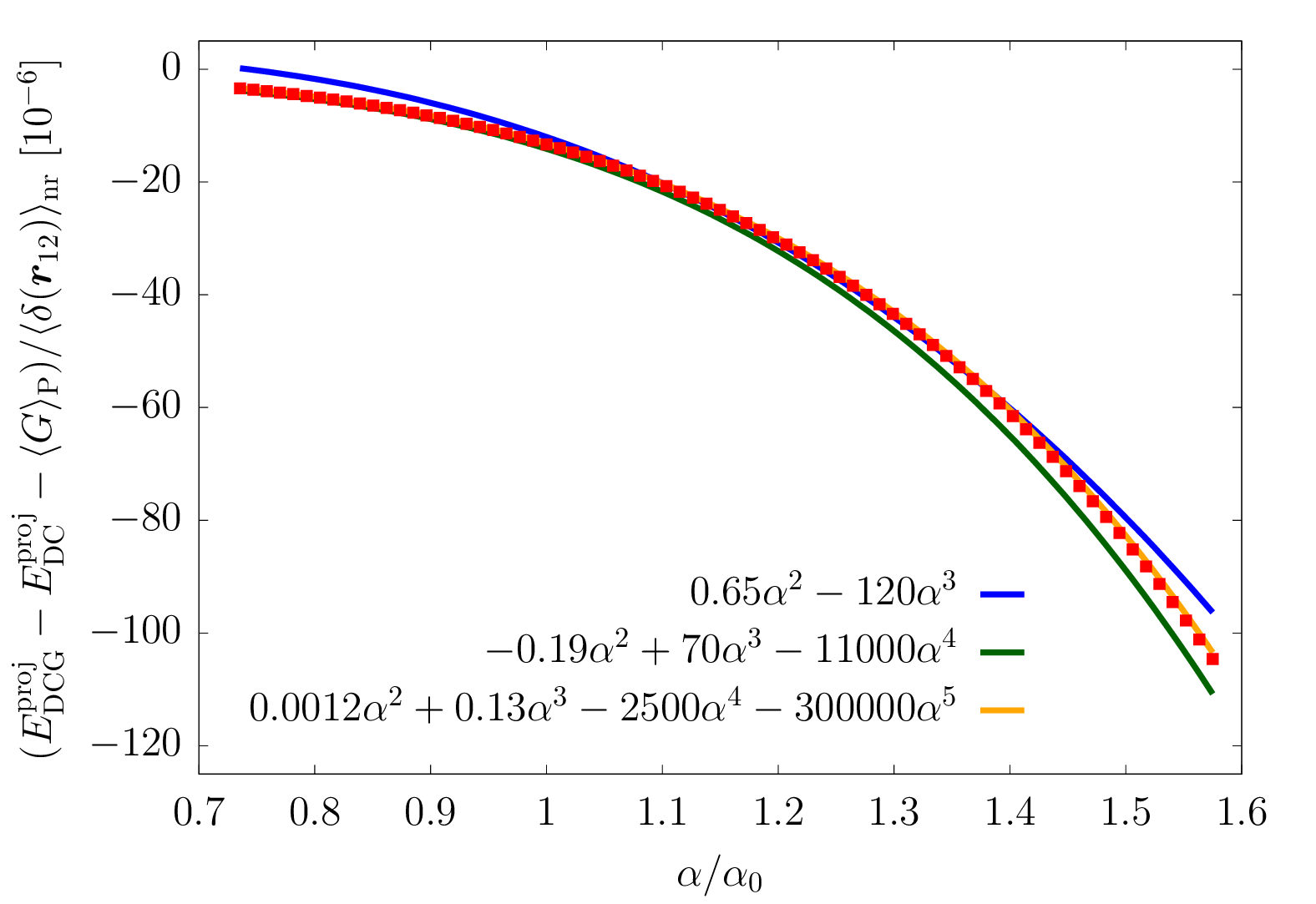} \\
    \caption{%
      Dependence of the no-pair Dirac--Coulomb--Gaunt (DCG) energy on the value of the $\alpha$ coupling constant of the electromagnetic interaction for the example of the ground state of the helium atom. 
      Hartree atomic units are used and $\alpha_0$ labels $1/137.$035~999~084 \cite{codata18}. 
      The $\alpha$ dependence of the no-pair Dirac--Coulomb energy, $E_\DC^\proj$, was shown in Fig.~3 of Paper~I.
      The leading-order $\alpha^2$ perturbative value for the Gaunt correction was obtained as $\langle G \rangle_\text{P}$ (using the Pauli approximation and our sixteen-component Gaunt implementation). 
      The $\langle \delta(\bos{r}_{12})\rangle_\text{nr}$ value was taken from Ref.~\cite{Dr06}. 
      %
      \label{fig:DCGalphafit}
    }
\end{figure}

We have repeated all computations using only the Gaunt term, Eq.~(\ref{eq:gaunt}), \emph{i.e.,} solved the no-pair Dirac--Coulomb--Gaunt equation and observed a similar behaviour (Fig.~\ref{fig:DCGalphafit}): 
(a) relatively slow convergence with respect to the basis set size;
(b) failure to find a stable fit of $\sum_{n=2}^m g_n \alpha^n$ polynomials by changing $m=3$ to 4. 

Since the Dirac--Coulomb--Gaunt and Dirac--Coulomb--Breit wave functions have different coalescence properties \cite{LiShLi12}, it is unlikely that the similar erratic behaviour of the DCG and DCB results (Figs.~\ref{fig:DCBalphafit} and \ref{fig:DCGalphafit}) is caused by the inexact representation of the coalescence features by the ECG basis set. 

For further progress along these lines, it will be necessary to 
(a) improve the convergence of the no-pair DCB (and DCG) energies by an order of magnitude;
(b) compute pair-corrections to the no-pair energies; and 
(c) test the kinetic balance condition to be able to rule out any deviation caused by an unnoticed `prolapse'. 
We have not noticed any major sign of prolapse during our no-pair DC computations (Paper~I), but it cannot be entirely excluded since the `restricted' kinetic balance, Eq.~(\ref{eq:kb}), is only an approximation to the `atomic balance' \cite{LeSe10} that would ensure a rigorous variational property for the computation of the non-interacting states used to define the positive-energy projector. It would be prohibitively difficult to use the atomic balance in this work, but we consider testing other types of (approximate) kinetic balance conditions \cite{shabaevDualKineticBalance2004,SiMaRe15}.

\clearpage
\section{Summary and conclusion}
In summary, the implementation of the Breit interaction operator has been reported in the explicitly correlated no-pair Dirac--Coulomb framework described in Paper~I \cite{JeFeMa21}. Both the variational and the perturbative inclusion of the Breit interaction was considered.
Numerical results have been reported for the 
helium atom and helium-like ions with small nuclear charge numbers as well as 
for the ground electronic state of the H$_2$, HeH$^+$, and H$_3^+$ molecular systems near their equilibrium configuration.

The numerical results are compared with the leading-order, $\mathcal{O}(\alpha^2)$, relativistic energies and the relevant, leading-order, $\mathcal{O}(\alpha^3)$, non-radiative quantum electrodynamics corrections that are available from non-relativistic quantum electrodynamics derivations in the literature. Further work is necessary to clarify the origin of the deviation of the variational and the perturbative relativistic treatments.

\vspace{0.5cm}
\section*{Data availability statement}
The data that support findings of this study is included in the paper or in the Supplementary Material.

\vspace{0.5cm}
\section*{Supplementary Material}
The supplementary material contains 
(a) matrix elements for the Breit operator;
(b) convergence tables;
(c) collection of expectation values for the evaluation of the perturbative formulae.

\vspace{0.5cm}
\begin{acknowledgments}
\noindent Financial support of the European Research Council through a Starting Grant (No.~851421) is gratefully acknowledged. DF thanks a doctoral scholarship from the ÚNKP-21-3 New National Excellence Program of the Ministry for Innovation and Technology from the source of the National Research, Development and Innovation Fund (ÚNKP-21-3-II-ELTE-41). %
We thank the Reviewers for their thoughtful comments that helped us to improve this work (Paper I \& II).
\end{acknowledgments}

\clearpage
%

%
\clearpage

\setcounter{section}{0}
\renewcommand{\thesection}{S\arabic{section}}
\setcounter{subsection}{0}
\renewcommand{\thesubsection}{S\arabic{section}.\arabic{subsection}}

\setcounter{equation}{0}
\renewcommand{\theequation}{S\arabic{equation}}

\setcounter{table}{0}
\renewcommand{\thetable}{S\arabic{table}}

\setcounter{figure}{0}
\renewcommand{\thefigure}{S\arabic{figure}}

~\\[0.cm]
\begin{center}
\begin{minipage}{0.8\linewidth}
\centering
\textbf{Supplementary Material} \\[0.25cm]

\textbf{On the Breit interaction in an explicitly correlated variational Dirac--Coulomb framework}
\end{minipage}
~\\[0.5cm]
\begin{minipage}{0.6\linewidth}
\centering

D\'avid Ferenc,$^1$ P\'eter Jeszenszki,$^1$ and Edit M\'atyus$^{1,\ast}$ \\[0.15cm]

$^1$~\emph{ELTE, Eötvös Loránd University, Institute of Chemistry, 
Pázmány Péter sétány 1/A, Budapest, H-1117, Hungary} \\[0.15cm]
$^\ast$ edit.matyus@ttk.elte.hu \\
\end{minipage}
~\\[0.15cm]
(Dated: January 4, 2022)
\end{center}

~\\[1cm]
\begin{center}
\begin{minipage}{0.9\linewidth}
\noindent %
Contents: \\
S1. Matrix elements of the Breit operator \\
S2. Convergence tables \\
S3. Collection of expectation values for the evaluation of the perturbative formulae \\
\end{minipage}
\end{center}

\newpage
\section{Matrix elements of the Breit operator}
\noindent The Breit operator is
\begin{align}
    B(1,2)=-\frac{q_1q_2}{2}\left[\frac{\balpha_1\balpha_2}{r_{12}}+\frac{(\bos{r}_{12}\cdot\balpha_1)(\bos{r}_{12}\cdot\balpha_2)}{r_{12}^3}\right]
\end{align}
or equivalently
\begin{align}
    B(1,2)=-q_1q_2\left[\frac{1}{r_{12}}\balpha_1\balpha_2+\frac{1}{2}\lbrace(\balpha_1\cdot\grad_1)(\balpha_2\cdot\grad_2)r_{12}\rbrace\right] \, ,
    \label{eq:Breitwork}
\end{align}
where the derivatives act only on $r_{12}$. The first term is the Gaunt interaction
\begin{align}
    G(1,2)=-\frac{q_1q_2}{r_{ij}}\balpha_1\balpha_2=\begin{pmatrix}
     0 & 0 & 0 & G \\
     0 & 0 & G & 0 \\
     0 & G & 0 & 0 \\
     G & 0 & 0 & 0 \\
    \end{pmatrix} \, .
\end{align}
The $4\times 4$ blocks of this 16-dimensional matrix are
\begin{align}
    G=-\frac{q_1q_2}{r_{12}}\bsigma_1\cdot\bsigma_2=-\frac{q_1q_2}{r_{12}}\begin{pmatrix*}[r]
     1 & 0 & 0 & 0 \\
     0 &-1 & 2 & 0 \\
     0 & 2 &-1 & 0 \\
     0 & 0 & 0 & 1
    \end{pmatrix*} \, .
\end{align}
After the kinetic balance transformation we obtain the following blocks 
\begin{align}
    G_1&=\frac{G(\bsigma_1\cdot\bp_1)(\bsigma_2\cdot\bp_2)}{4c^2m_1m_2} \label{G1matrix} \\
    G_2&=\frac{(\bsigma_2\cdot\bp_2)G(\bsigma_1\cdot\bp_1)}{4c^2m_1m_2}\\
    G_3&=\frac{(\bsigma_1\cdot\bp_1)G(\bsigma_2\cdot\bp_2)}{4c^2m_1m_2}\\
    G_4&=\frac{(\bsigma_2\cdot\bp_2)(\bsigma_1\cdot\bp_1)G}{4c^2m_1m_2} \ .
\end{align}
The necessary matrix elements are evaluated as
\begin{align}
    &\matrixel{\Theta_\mu}{\frac{1}{r_{ij}}p_{1_k}p_{2_l}}{\Theta_\nu}=-\intdr \frac{1}{r_{ij}} \Theta_\mu \partial_{1_k}\partial_{2_l} \Theta_\nu \nonumber \\
    &=2 (\bA_\nu)_{12}\delta_{kl} \intdr \frac{1}{r_{ij}} \Theta_\mu\Theta_\nu-4 \intdr \frac{1}{r_{ij}} (\br-\bs_\nu)\bA_\nu \left[ \bE_{12} \otimes \bos{\varepsilon}_{kl} \right]\bA_\nu  (\br-\bs_\nu)\Theta_\mu\Theta_\nu \, ,
\end{align}
where the resulting terms are obtained using standard methods for Gaussian integrals. The matrices $\bE_{12}$ and $\varepsilon_{kl}$ are of dimension $\np \times \np$ and $3\times 3$, respectively, and all their elements equal zero, except for the $(1,2)$ and $(k,l)$ element, which is one.

The blocks from the second term in Eq.~\eqref{eq:Breitwork} after the kinetic balance transformation are
\begin{align}
    B_1&=\frac{B(\bsigma_1\cdot\bp_1)(\bsigma_2\cdot\bp_2)}{4c^2m_1m_2} \label{B1op}\\
    B_2&=\frac{(\bsigma_2\cdot\bp_2)B(\bsigma_1\cdot\bp_1)}{4c^2m_1m_2}\\
    B_3&=\frac{(\bsigma_1\cdot\bp_1)B(\bsigma_2\cdot\bp_2)}{4c^2m_1m_2}\\
    B_4&=\frac{(\bsigma_2\cdot\bp_2)(\bsigma_1\cdot\bp_1)B}{4c^2m_1m_2} \ ,
\end{align}
where
\begin{align}
    B=\frac{1}{2}\sum_{i=1}^3\sum_{j=1}^3\sigma_{1_i} \sigma_{2_j} \lbrace \grad_{1_i}\grad_{2_j}r_{12} \rbrace \, .
\end{align}
For the sake of brevity, we introduce the notation
\begin{align}
    \mathcal{A}_{ij}=\frac{\sigma_{1_i} \sigma_{2_j}}{8c^2m_1m_2} \, .
\end{align}
Then, we write out the differential operators explicitly and act once on ${r_{12}}$,
\begin{align}
    B_1&=\sum_{i,j,k,l}\mathcal{A}_{ij} \sigma_{1_k}\sigma_{2_l} \left(\partial_{1_i}\partial_{2_j} r_{12}\right) p_{1_k} p_{2_l}=
    -\sum_{i,j,k,l}\mathcal{A}_{ij} \sigma_{1_k}\sigma_{2_l} \left(\partial_{1_i}\partial_{2_j} r_{12}\right) \partial_{1_k} \partial_{2_l}  \nonumber \\
    &=-\sum_{i,j,k,l}\mathcal{A}_{ij} \sigma_{1_k}\sigma_{2_l} \left(\partial_{2_j} \frac{r_{12_i}}{r_{12}}\right) \partial_{1_k} \partial_{2_l} 
\end{align}
\begin{align}
    B_2&=\sum_{i,j,k,l}\sigma_{2_k}\mathcal{A}_{ij} \sigma_{1_l} p_{2_k} \left(\partial_{1_i}\partial_{2_j} r_{12}\right) p_{1_l} =
    -\sum_{i,j,k,l}\sigma_{2_k}\mathcal{A}_{ij} \sigma_{1_l} \partial_{2_k} \left(\partial_{1_i}\partial_{2_j} r_{12}\right) \partial_{1_l}  \nonumber \\
    &=\sum_{i,j,k,l}\sigma_{2_k}\mathcal{A}_{ij} \sigma_{1_l} \partial_{2_k} \left(\partial_{1_i}\frac{r_{12_j}}{ r_{12}}\right) \partial_{1_l}  \, .
\end{align}
The corresponding matrix elements with the ECG basis functions, $\Theta_\mu$ and $\Theta_\nu$, are
\begin{align}
    \matrixel{\Theta_\mu}{B_1}{\Theta_\nu} &= -\sum_{i,j,k,l}\mathcal{A}_{ij} \sigma_{1_k}\sigma_{2_l} \matrixel{\Theta_\mu}{\left(\partial_{2_j} \frac{r_{12_i}}{r_{12}}\right) \partial_{1_k} \partial_{2_l} }{\Theta_\nu} \nonumber \\
    &=\sum_{i,j,k,l}\mathcal{A}_{ij} \sigma_{1_k}\sigma_{2_l} \left[ \matrixel{\partial_{2_j} \Theta_\mu}{\frac{r_{12_i}}{r_{12}} }{ \partial_{1_k} \partial_{2_l} \Theta_\nu}+\matrixel{ \Theta_\mu}{ \frac{r_{12_i}}{r_{12}} }{\partial_{2_j} \partial_{1_k} \partial_{2_l}\Theta_\nu} \right]
\end{align}
and
\begin{align}
    \matrixel{\Theta_\mu}{B_2}{\Theta_\nu} &= \sum_{i,j,k,l}\sigma_{2_k}\mathcal{A}_{ij} \sigma_{1_l} \matrixel{\Theta_\mu}{\partial_{2_k} \left(\partial_{1_i}\frac{r_{12_j}}{ r_{12}}\right) \partial_{1_l}}{\Theta_\nu} \nonumber \\
    &=-\sum_{i,j,k,l}\sigma_{2_k}\mathcal{A}_{ij} \sigma_{1_l} \matrixel{ \partial_{2_k}  \Theta_\mu}{\left(\partial_{1_i}\frac{r_{12_j}}{ r_{12}}\right) \partial_{1_l}}{\Theta_\nu}\nonumber \\
    &=\sum_{i,j,k,l}\sigma_{2_k}\mathcal{A}_{ij} \sigma_{1_l} \left[  \matrixel{\partial_{1_i} \partial_{2_k}  \Theta_\mu}{\frac{r_{12_j}}{ r_{12}} }{\partial_{1_l}\Theta_\nu} + \matrixel{ \partial_{2_k}  \Theta_K}{\frac{r_{12_j}}{ r_{12}} }{\partial_{1_i}\partial_{1_l}\Theta_\nu}\right] \; .
\end{align}
Since the Hamiltonian matrix is Hermitian, these are the only matrix elements that we need to calculate, and only two genuinely distinct types of integrals appear in the formulae:

\begin{align}
    &I_1(\mu,\nu,a,b,c,i,j,k,l) = \intdr \frac{r_{12_i}}{r_{12}}  \left(\partial_{a_j} \Theta_\mu \right) \left(\partial_{b_k} \partial_{c_l} \Theta_\nu  \right)  \nonumber \\ 
    &=4 (\bA_\nu)_{bc}\delta_{kl}  \intdr \frac{1}{r_{12}}  (\br-\bs_\mu)^T \bA_\mu \be_{a,j} \br^T\be_{12,i} \Theta_\mu\Theta_\nu \nonumber \\
    &-8 \intdr \frac{1}{r_{12}} (\br-\bs_\mu)^T \bA_\mu \left[ \be_{a,j} \otimes \be_{12,i}  \right]  \br \  (\br-\bs_\nu)^T \bA_\nu \left[\be_{b,k} \otimes  \be_{c,l} \right]\bA_\nu (\br-\bs_\nu)  \Theta_\mu \Theta_\nu
    \label{eq:funI1}
\end{align}
and
\begin{align}
    &I_2(\mu,\nu,a,b,c,i,j,k,l) = \intdr   \frac{r_{12_i}}{r_{12}}  \Theta_\mu \left( \partial_{a_j} \partial_{b_k} \partial_{c_l}\Theta_\nu \right)  \nonumber \\
    &=4(\bA_\nu)_{ab}\delta_{jk} \intdr \frac{1}{r_{12}} \br^T\be_{12,i}  (\br-\bs_\nu)^T\bA_\nu\be_{c,l} \Theta_\mu\Theta_\nu \nonumber \\ &+4(\bA_\nu)_{bc}\delta_{kl}\intdr \frac{1}{r_{12}} \br^T\be_{12,i} (\br-\bs_\nu)^T\bA_\nu\be_{a,j} \Theta_\mu\Theta_\nu \nonumber \\
    &+4(\bA_\nu)_{ac}\delta_{jl}\intdr \frac{1}{r_{12}} \br^T\be_{12,i} (\br-\bs_\nu)^T\bA_\nu\be_{b,k} \Theta_\mu\Theta_\nu \nonumber \\
    &-8\intdr \frac{1}{r_{12}}  (\br-\bs_\nu)^T\bA_\nu\left[\be_{b,k} \otimes \be_{12,i} \right] \br (\br-\bs_\nu)^T\bA_\nu\left[\be_{c,l} \otimes \be_{a,j} \right]\bA_\nu (\br-\bs_\nu) \Theta_\mu \Theta_\nu \; ,
    \label{eq:funI2}
\end{align}
where we used the following notation: general particle indices are $a,b \in \{1,\dots,\np\}$, the Cartesian components are labelled with $i,j,k,l\in\{x,y,z\}$, the basis functions are labelled by Greek letters $\mu,\nu$. The vector $\bos{e}_{a,i}$ is the unit vector with zero elements everywhere, except for it's $a,i$ element, which is $1$. $\bos{e}_{ab,i}$ is a vector with zero elements everywhere, except for it's $a,i$ element, which is $1$, and it's $b,i$ element, which is $-1$, \emph{i.e.,} $\be^T_{ab,i}\bos{v}=(v_a-v_b)_i$. 

The matrix elements can be written using the $I_1$ and $I_2$ functions, Eqs.~(\ref{eq:funI1})--(\ref{eq:funI2}):
\begin{align}
    \matrixel{\Theta_\mu}{B_1}{\Theta_\nu} = 
    &\sum_{i,j,k,l}\mathcal{A}_{ij} \sigma_{1_k}\sigma_{2_l} \left[ \matrixel{\partial_{2_j} \Theta_\mu}{\frac{r_{12_i}}{r_{12}} }{ \partial_{1_k} \partial_{2_l} \Theta_\nu}+\matrixel{ \Theta_\mu}{ \frac{r_{12_i}}{r_{12}} }{\partial_{2_j} \partial_{1_k} \partial_{2_l}\Theta_\nu} \right]  \nonumber \\
    =&\sum_{i,j,k,l}\mathcal{A}_{ij} \sigma_{1_k}\sigma_{2_l} \left[ I_1(\mu,\nu,2,1,2,i,j,k,l)+I_2(\mu,\nu,2,1,2,i,j,k,l) \right]  \\
    \matrixel{\Theta_\mu}{B_2}{\Theta_\nu} = 
    &\sum_{i,j,k,l}\sigma_{2_k}\mathcal{A}_{ij} \sigma_{1_l} \left[  \matrixel{\partial_{1_i} \partial_{2_k}  \Theta_\mu}{\frac{r_{12_j}}{ r_{12}} }{\partial_{1_l}\Theta_\nu} + \matrixel{ \partial_{2_k}  \Theta_\mu}{\frac{r_{12_j}}{ r_{12}} }{\partial_{1_i}\partial_{1_l}\Theta_\nu}\right]  \nonumber \\
    =&\sum_{i,j,k,l}\sigma_{2_k}\mathcal{A}_{ij} \sigma_{1_l} \left[ I_1(\nu,\mu,1,1,2,j,l,i,k) + I_1(\mu,\nu,2,1,1,j,k,i,l) \right] \\
    \matrixel{\Theta_\mu}{B_3}{\Theta_\nu} = 
    &\sum_{i,j,k,l}\sigma_{1_k}\mathcal{A}_{ij} \sigma_{2_l} \left[ \matrixel{\partial_{1_k}\Theta_\mu}{\frac{r_{12_j}}{ r_{12}}}{\partial_{1_i}\partial_{2_l}\Theta_\nu} + \matrixel{\partial_{1_k}\partial_{1_i}\Theta_\mu}{\frac{r_{12_j}}{ r_{12}}}{\partial_{2_l}\Theta_\nu} \right]  \nonumber \\
    =&\sum_{i,j,k,l}\sigma_{1_k}\mathcal{A}_{ij} \sigma_{2_l}  \left[ I_1(\mu,\nu,1,1,2,j,k,i,l) + I_1(\nu,\mu,2,1,1,j,l,k,i) \right] \\
    \matrixel{\Theta_\mu}{B_4}{\Theta_\nu} = 
    &\sum_{i,j,k,l} \sigma_{2_k}\sigma_{1_l}\mathcal{A}_{ij}  \left[ \matrixel{\partial_{2_k}\partial_{1_l}\Theta_\mu}{\frac{r_{12_i}}{ r_{12}}}{\partial_{2_j} \Theta_\nu} + \matrixel{\partial_{2_j}\partial_{2_k}\partial_{1_l}\Theta_\mu}{\frac{r_{12_i}}{ r_{12}}}{ \Theta_\nu} \right] \nonumber \\
    =&\sum_{i,j,k,l} \sigma_{2_k}\sigma_{1_l}\mathcal{A}_{ij}  \left[ I_1(\nu,\mu,2,2,1,i,j,k,l)+I_2(\nu,\mu,2,2,1,i,j,k,l) \right] 
\end{align}

\clearpage

\section{Convergence tables}
\begin{table}[h]
  \caption{%
    Convergence of 
    the no-pair Dirac--Coulomb and Dirac--Coulomb--Breit energies, in $\Eh$, for the 
    ground electronic state of the H$_2$ molecule ($R_\text{pp}=1.4$~bohr) 
    with respect to the $N_\text{b}$ number of explicitly correlated Gaussian (ECG) basis functions.
    \label{tab:h2}
  }
  \begin{tabular}{@{}l rr rr@{}}
    \hline\hline\\[-0.35cm]
    \multicolumn{1}{c}{$\nb$} &	
    \multicolumn{1}{c}{$E_\DC^\proj$}  &
    \multicolumn{1}{c}{$E_\DC^\proj+\langle B \rangle_\DC$} &
    \multicolumn{1}{c}{$E_\DC^\proj+\mathcal{P}^{(2)}\lbrace B\rbrace$} &
    \multicolumn{1}{c}{$E_\DCB^\proj$} \\
    \cline{1-5}\\[-0.35cm]
    128	&	$-$1.174 489 583  	&	$-$1.174 486 463  	&	$-$1.174 486 474  	&	$-$1.174 486 474	\\
    256	&	$-$1.174 489 738  	&	$-$1.174 486 640  	&	$-$1.174 486 669  	&	$-$1.174 486 669	\\
    512	&	$-$1.174 489 753  	&	$-$1.174 486 660  	&	$-$1.174 486 701  	&	$-$1.174 486 701	\\
    700	&	$-$1.174 489 754  	&	$-$1.174 486 663  	&	$-$1.174 486 711  	&	$-$1.174 486 711	\\
    800	&	$-$1.174 489 754  	&	$-$1.174 486 663  	&	$-$1.174 486 713  	&	$-$1.174 486 712	\\
    1000&	$-$1.174 489 754  	&	$-$1.174 486 664  	&	$-$1.174 486 718  	&	$-$1.174 486 717	\\
    1200&	$-$1.174 489 754  	&	$-$1.174 486 665  	&	$-$1.174 486 722  	&	$-$1.174 486 721	\\
    \cline{1-5}\\[-0.35cm]
    $E-E^{(2)}_\text{DCB}$ $^\text{a}$  &  
       	&	0.000 000 002	&	$-$0.000 000 055	&	$-$0.000 000 054	\\
    \hline\hline
    \end{tabular}
  \begin{flushleft}
    $^\text{a}$ %
    $E^{(2)}_\text{DCB}=-$1.174 486 667~$\Eh$\ \cite{PuKoPa17,PuKoCzPa16}. 
  \end{flushleft}    
\end{table}

\begin{table}[h]
  \caption{%
    Convergence of the 
    no-pair Dirac--Coulomb and Dirac--Coulomb--Breit energies, in $\Eh$, for the 
    ground electronic state of the H$_3^+$ molecular ion ($R_\text{pp}=1.65$~bohr)
    with respect to the $N_\text{b}$ number of explicitly correlated Gaussian (ECG) basis functions.
    The basis set size corresponds to the use of $D_{3\text{h}}$
    point-group symmetry in the computations.    
    \label{tab:h3p}
  }
  \begin{tabular}{@{}l rr rr@{}}
    \hline\hline\\[-0.35cm]
    \multicolumn{1}{l}{$\nb$} &	
    \multicolumn{1}{c}{$E_\DC^\proj$}  &
    \multicolumn{1}{c}{$E_\DC^\proj+\langle B \rangle$} &
    \multicolumn{1}{c}{$E_\DC^\proj+\mathcal{P}^{(2)}\lbrace B\rbrace$} &
    \multicolumn{1}{c}{$E_\DCB^\proj$} \\
    \cline{1-5}\\[-0.35cm]
    100    & $-$1.343 850 149 & $-$1.343 847 001 & $-$1.343 847 014
 & $-$1.343 847 014 \\
    200    & $-$1.343 850 507 & $-$1.343 847 379 & $-$1.343 847 404 & $-$1.343 847 404 \\
    300    & $-$1.343 850 524 & $-$1.343 847 409 & $-$1.343 847 462 & $-$1.343 847 462 \\
    400    & $-$1.343 850 526 & $-$1.343 847 414 & $-$1.343 847 485 & $-$1.343 847 484 \\
    500    & $-$1.343 850 527 & $-$1.343 847 416 & $-$1.343 847 497 & $-$1.343 847 496 \\
    600    & $-$1.343 850 527 & $-$1.343 847 416 & $-$1.343 847 499 & $-$1.343 847 498 \\    
    \cline{1-5}\\[-0.35cm]
    $E-E^{(2)}_\text{DCB}$ $^\text{a}$  &  
    &	 0.000 000 000	&	$-$0.000 000 083
	&	$-$0.000 000 082	\\
    \hline\hline
    \end{tabular}
  \begin{flushleft}
    $^\text{a}$ %
    $E^{(2)}_\text{DCB}=-$1.343 847 416~$\Eh$\ \cite{JeIrFeMa21}.
  \end{flushleft}    
\end{table}

\begin{table}[h]
  \caption{%
    Convergence of the
    no-pair Dirac--Coulomb and Dirac--Coulomb--Breit energies, in $\Eh$, for the 
    ground electronic state of the HeH$^+$ molecular ion ($R_\text{pp}=1.46$~bohr)
    with respect to the $N_\text{b}$ number of explicitly correlated Gaussian (ECG) basis functions.
    \label{tab:hehp}
  }
  \begin{tabular}{@{}l rr rr@{}}
    \hline\hline\\[-0.35cm]
    \multicolumn{1}{c}{$\nb$} &	
    \multicolumn{1}{c}{$E_\DC^\proj$}  &
    \multicolumn{1}{c}{$E_\DC^\proj+\langle B \rangle$} &
    \multicolumn{1}{c}{$E_\DC^\proj+\mathcal{P}^{(2)}\lbrace B\rbrace$} &
    \multicolumn{1}{c}{$E_\DCB^\proj$} \\
    \cline{1-5}\\[-0.35cm]
    400	 &	$-$2.978 834 584	&	$-$2.978 808 074	&	$-$2.978 808 255	&	$-$2.978 808 249	\\
    600	 &	$-$2.978 834 630	&	$-$2.978 808 174	&	$-$2.978 808 555	&	$-$2.978 808 552	\\
    800	 &	$-$2.978 834 634	&	$-$2.978 808 191	&	$-$2.978 808 686	&	$-$2.978 808 682	\\
    1000 &	$-$2.978 834 635	&	$-$2.978 808 196	&	$-$2.978 808 741	&	$-$2.978 808 699	\\
    1200 &	$-$2.978 834 635	&	$-$2.978 808 200	&	$-$2.978 808 780	&	$-$2.978 808 774	\\
    \cline{1-5}\\[-0.35cm]
    $E-E^{(2)}_\text{DCB}$ $^\text{a}$  &  
    &	$-$0.000 000 020	&	$-$0.000 000 600	&	$-$0.000 000 594	\\
    \hline\hline
    \end{tabular}
  \begin{flushleft}
    $^\text{a}$ %
    $E^{(2)}_\text{DCB}=-$2.978 808 180~$\Eh$\ \cite{JeFeMa21}.
  \end{flushleft}    
\end{table}

\begin{table}[h]
  \caption{%
    Convergence of the  
    no-pair Dirac--Coulomb and Dirac--Coulomb--Breit energies, in $\Eh$, for the 1~and~2~$^1$S$_0$ states of the helium atom 
    with respect to the $N_\text{b}$ number of explicitly correlated Gaussian (ECG) basis functions.
    \label{tab:he}
  }
  \begin{tabular}{@{}c l rr rr@{}}
    \hline\hline\\[-0.35cm]
    &\multicolumn{1}{c}{$\nb$} &	
    \multicolumn{1}{c}{$E_\DC^\proj$}  &
    \multicolumn{1}{c}{$E_\DC^\proj+\langle B \rangle$} &
    \multicolumn{1}{c}{$E_\DC^\proj+\mathcal{P}^{(2)}\lbrace B\rbrace$} &
    \multicolumn{1}{c}{$E_\DCB^\proj$} 
    \\
    \cline{2-6}\\[-0.35cm]
    &100	&	$-$2.903 856 311	& $-$2.903 827 961	& $-$2.903 828 083 & $-$2.903 828 081 \\
    &200	&	$-$2.903 856 622	& $-$2.903 828 281	& $-$2.903 828 614 & $-$2.903 828 609 \\[-0.3cm]
    \raisebox{0.3cm}{1~$^1$S$_0$}&300	&	$-$2.903 856 631	& $-$2.903 828 328	& $-$2.903 828 976 & $-$2.903 828 968 \\
    &400	&	$-$2.903 856 631	& $-$2.903 828 333	& $-$2.903 829 031 & $-$2.903 829 023 \\
    \hline\\[-0.35cm]
    &$E-E^{(2)}_\text{DCB}$ $^\text{a}$  &  
      &$-$0.000 000 022  &  $-$0.000 000 720    &	$-$0.000 000 712 \\[0.05cm]
    \hline\\[-0.35cm]
    &100	&	$-$2.146 084 035	&	$-$2.146 081 591	&	$-$2.146 081 594	&	$-$2.146 081 595	\\
    &200	&	$-$2.146 084 756	&	$-$2.146 082 344	&	$-$2.146 082 386	&	$-$2.146 082 386	\\[-0.30cm]
    \raisebox{0.3cm}{2~$^1$S$_0$} &300	&	$-$2.146 084 789	&	$-$2.146 082 377	&	$-$2.146 082 421	&	$-$2.146 082 421	\\
    &400	&	$-$2.146 084 791	&	$-$2.146 082 379	&	$-$2.146 082 425	&	$-$2.146 082 424	\\
    \hline\\[-0.35cm] 
    &$E-E^{(2)}_\text{DCB}$ $^\text{b}$  &  
     &	$-$0.000 000 011	&	$-$0.000 000 056	&	$-$0.000 000 056	\\
    \hline\hline
    \end{tabular}
  \begin{flushleft}
    $^\text{a}$ %
    $E^{(2)}_\text{DCB}=-$2.903 828 311~$\Eh$\ \cite{Dr06}. \\
    $^\text{b}$ %
    $E^{(2)}_\text{DCB}=-$2.146 082 368~$\Eh$\ \cite{Dr88}. \\
  \end{flushleft}    
\end{table}

\clearpage
\section{Collection of expectation values used for the evaluation of the perturbative formulae}
\begin{table}[h]
  \caption{%
    Reference values for the non-relativistic energy and expectation values of relativistic correction terms with the non-relativistic wave function. In this table $\langle O\rangle$ means $\langle O\rangle_\text{nr}$. See also Sec.~III.B of the manuscript.
    \label{tab:som-pt-lit}
  }
  \centering
  \begin{tabular}{@{}l rr@{\ \ \ } rrr c@{}}
    \hline\hline\\[-0.35cm]
    	&
    \multicolumn{1}{c}{$E_\text{nr}$}	&	
    \multicolumn{1}{c}{$-\frac{1}{8}\sum_{i} \langle (\bos{p}_i^2)^2 \rangle$}	&	
    \multicolumn{1}{c}{$\sum_{i,a} Z_a \langle \delta(\bos{r}_{ia})\rangle$}	&	
    \multicolumn{1}{c}{$\langle \delta(\bos{r}_{12})\rangle$} &
    \multicolumn{1}{c}{$\langle H_\text{OO}\rangle$} & 
    \multicolumn{1}{c}{Ref.} \\
    \hline\\[-0.35cm]
    H$^-$	    &	$-$0.527 751 017	&	$-$0.615 640	&	0.329 106	&	0.002 738	&	$-$0.008 875	& \cite{Dr06} \\
    He      	&	$-$2.903 724 377	&	$-$13.522 017	&	7.241 717	&	0.106 345	&	$-$0.139 095	& \cite{Dr06} \\
    Li$^+$	    &	$-$7.279 913 413	&	$-$77.636 788	&	41.112 057	&	0.533 723	&	$-$0.427 992	& \cite{Dr06} \\
    Be$^{2+}$	&	$-$13.655 566 238	&	$-$261.819 623	&	137.585 380	&	1.522 895	&	$-$0.878 769	& \cite{Dr06} \\
    \hline \\[-0.35cm]													
    H$_2$	&	$-$1.174 475 714	&	$-$1.654 745	&	0.919 336	&	0.016 743	&	$-$0.047 634	&	\cite{PuKoPa17} \\
    H$_3^+$	&	$-$1.343 835 625	&	$-$1.933 424	&	1.089 655	&	0.018 335	&	$-$0.057 218	&	\cite{JeIrFeMa21} \\
    HeH$^+$	&	$-$2.978 706 599	&	$-$13.419 287	&	7.216 253	&	0.101 122	&	$-$0.141 242	& \cite{JeFeMa21} \\
    \hline\hline\\[-0.35cm]
    	&	
    \multicolumn{1}{c}{$\langle H_\DC^{(2)} \rangle$} & 
    \multicolumn{1}{c}{$\langle H_\text{B}^{(2)} \rangle$} & 
    \multicolumn{1}{c}{$\langle H_\DC^{(2)}+H_\text{B}^{(2)} \rangle$} & 	
    \multicolumn{1}{c}{$E^{(2)}_\DC$} &	
    \multicolumn{1}{c}{$E^{(2)}_\DCB$}	&	
    \multicolumn{1}{c}{Ref.}	\\
    \hline\\[-0.35cm]
    H$^-$	   &	$-$0.107 283	&	0.008 328	&	$-$0.098 955	&	$-$0.527 756 730	&	$-$0.527 756 286	& \cite{Dr06}		\\
    He	       &	$-$2.480 848	&	0.529 093	&	$-$1.951 755	&	$-$2.903 856 486	&	$-$2.903 828 311	& \cite{Dr06}\\
    Li$^+$	   &	$-$14.734 859	&	2.925 486	&	$-$11.809 373	&	$-$7.280 698 064	&	$-$7.280 542 278	& \cite{Dr06} \\
    Be$^{2+}$  &	$-$50.485 330	&	8.689 865	&	$-$41.795 465	&	$-$13.658 254 651	&	$-$13.657 791 904	& \cite{Dr06}\\
    \hline\\[-0.35cm]
    H$_2$	&  $-$0.263 255	&	0.057 567	&	$-$0.205 689	&	$-$1.174 489 733	&	$-$1.174 486 667	& \cite{PuKoPa17}	\\
    H$_3^+$	&  $-$0.279 399	&	0.057 983	&	$-$0.221 416	&	$-$1.343 850 503	&	$-$1.343 847 416	& \cite{JeIrFeMa21} \\
    HeH$^+$	&  $-$2.401 709	&	0.494 128	&	$-$1.907 581	&	$-$2.978 834 493	&	$-$2.978 808 180	& \cite{JeFeMa21} \\
    \hline\hline\\[-0.35cm]
  \end{tabular}
\end{table}

\begin{table}[h]
  \caption{%
    Non-relativistic energy and expectation values of relativistic correction terms obtained in `direct' (non-regularized) computation with the  non-relativistic wave function using the largest basis sets optimized in this work (for each system, separately) and used to define a spatial basis set for no-pair Dirac--Coulomb(--Breit) computations. In this table $\langle {O}\rangle$ means $\langle {O}\rangle_\text{nr}$. See also Sec.~III.B of the manuscript.
    \label{tab:som-pt-lit}
  }
  \centering
  \begin{tabular}{@{}l rr@{\ \ \ } rrr @{}}
    \hline\hline\\[-0.35cm]
    	&
    \multicolumn{1}{c}{$E_\text{nr}$}	&	
    \multicolumn{1}{c}{$-\frac{1}{8}\sum_{i} \langle (\bos{p}_i^2)^2 \rangle$}	&	
    \multicolumn{1}{c}{$\sum_{i,a} Z_a \langle \delta(\bos{r}_{ia})\rangle$}	&	
    \multicolumn{1}{c}{$\langle \delta(\bos{r}_{12})\rangle$} &
    \multicolumn{1}{c}{$\langle H_\text{OO}\rangle$} \\
    \hline\\[-0.35cm]
    H$^-$	    & $-$0.527 751 016	&	$-$0.615 332	&	0.328 923	&	0.002 743	&	$-$0.008 875	\\
    He      	& $-$2.903 724 377	&	$-$13.519 096	&	7.239 442	&	0.106 448	&	$-$0.139 095	\\
    Li$^+$	    & $-$7.279 913 410	&	$-$77.562 651	&	41.067 671	&	0.534 083	&	$-$0.427 992	\\
    Be$^{2+}$	& $-$13.655 566 234	&	$-$261.660 823	&	137.491 296	&	1.523 969	&	$-$0.878 771	\\
    \hline \\[-0.35cm]													
    H$_2$	& $-$1.174 475 714	&	$-$1.653 578	&	0.918 653	&	0.016 768	&	$-$0.047 635	\\
    H$_3^+$	& $-$1.343 835 625	&	$-$1.932 048	&	1.088 845	&	0.018 358	&	$-$0.057 218	\\
    HeH$^+$	& $-$2.978 706 599	&	$-$13.406 476	&	7.208 549	&	0.101 223	&	$-$0.141 242	\\
    \hline\hline\\[-0.35cm]
    	&	
    \multicolumn{1}{c}{$\langle H_\DC^{(2)} \rangle$} & 
    \multicolumn{1}{c}{$\langle H_\text{B}^{(2)} \rangle$} & 
    \multicolumn{1}{c}{$\langle H_\DC^{(2)}+H_\text{B}^{(2)} \rangle$} & 	
    \multicolumn{1}{c}{$E^{(2)}_\DC$} &	
    \multicolumn{1}{c}{$E^{(2)}_\DCB$}	\\
    \hline\\[-0.35cm]
    H$^-$ & $-$0.107 279	&	0.008 361	&	$-$0.098 918	&	$-$0.527 756 729	&	$-$0.527 756 284	\\
    He & $-$2.481 823	&	0.529 736	&	$-$1.952 087	&	$-$2.903 856 537	&	$-$2.903 828 328	\\
    Li$^+$ & $-$14.731 576	&	2.927 751	&	$-$11.803 825	&	$-$7.280 697 887	&	$-$7.280 541 980	\\
    Be$^{2+}$ & $-$50.477 690	&	8.696 608	&	$-$41.781 082	&	$-$13.658 254 239	&	$-$13.657 791 133	\\
    \hline\\[-0.35cm]
    H$_2$ &	$-$0.263 240	&	0.057 725	&	$-$0.205 516	&	$-$1.174 489 732	&	$-$1.174 486 658	\\
    H$_3^+$ & $-$0.279 367	&	0.058 126	&	$-$0.221 240	&	$-$1.343 850 502	&	$-$1.343 847 407	\\ 
    HeH$^+$ & $-$2.401 316	&	0.494 763	&	$-$1.906 553	&	$-$2.978 834 472	&	$-$2.978 808 125	\\
    \hline\hline\\[-0.35cm]
  \end{tabular}
\end{table}

%

\end{document}